\documentclass[]{agujournal2019}
\let\linenumbers\relax

\usepackage{url} 
\usepackage{lineno}
\usepackage{microtype} 
\usepackage{amsmath} 
\usepackage[finalnew]{trackchanges} %for better track changes. finalnew option will compile document with changes incorporated. inline to see the changes
\addeditor{AJC}
\usepackage{soul}
\linenumbers

\journalname{JGR: Atmospheres}
\begin{document}
\title{Towards real-time assessment of infrasound event detection capability using deep learning-based transmission loss estimation}

%%%%%%%%%%%%%%%%%%%%%%%%%%%%%%%%%%%%%%%%%%%%%%%
%  AUTHORS AND AFFILIATIONS
%%%%%%%%%%%%%%%%%%%%%%%%%%%%%%%%%%%%%%%%%%%%%%%
\authors{A. Janela Cameijo\affil{1,2}, A. Le Pichon\affil{1}, Y. Sklab\affil{3}, S. Arib\affil{4}, Q. Brissaud\affil{5}, S.P. Näsholm\affil{5,6}, C. Listowski\affil{1}, S. Aknine\affil{2}}

\affiliation{1}{CEA, DAM, DIF, F--91297 Arpajon, France}
\affiliation{2}{Université Lyon 1, LIRIS, Nautibus, Boulevard du 11 Novembre 1918, F--69100, Villeurbanne, France}
\affiliation{3}{Sorbonne Université, IRD, UMMISCO, 32 Avenue Henri Varagnat, F--93143, Bondy, France}
\affiliation{4}{CY Cergy Paris Université, Laboratoire Thema, CNRS UMR 8184, 33 Boulevard du Port, F--95011, Cergy-Pontoise, France}
\affiliation{5}{NORSAR, Solutions Department, Gunnar Randers vei 15, N--2007, Kjeller, Norway}
\affiliation{6}{University of Oslo, Department of Informatics, Problemveien 11, N--0316, Oslo, Norway}

\correspondingauthor{Alice Janela Cameijo}{alice.cameijo@cea.fr}

%%%%%%%%%%%%%%%%%%%%%%%%%%%%%%%%%%%%%%%%%%%%%%%
% KEY POINTS
%%%%%%%%%%%%%%%%%%%%%%%%%%%%%%%%%%%%%%%%%%%%%%%
\begin{keypoints}
\item We provide a convolutional recurrent neural network estimating in near real-time ground-level infrasound transmission loss.
\item The neural network exploits spatial and range-dependent features in atmospheric models, and makes predictions with an uncertainty estimate.
\item The network can be used as a tool for near real-time estimation of infrasound event detection capability at a global scale.
\end{keypoints}

%%%%%%%%%%%%%%%%%%%%%%%%%%%%%%%%%%%%%%%%%%%%%%%
%  ABSTRACT and PLAIN LANGUAGE SUMMARY
%%%%%%%%%%%%%%%%%%%%%%%%%%%%%%%%%%%%%%%%%%%%%%%
\begin{abstract}
Accurate modeling of infrasound transmission loss is essential for evaluating the performance of the International Monitoring System\change[AJC]{, which monitors compliance with the Comprehensive Nuclear-Test-Ban Treaty.}{, enabling the effective design and maintenance of infrasound stations to support compliance of the Comprehensive Nuclear-Test-Ban Treaty.} \add[AJC]{State-of-the-art} propagation modeling tools enable transmission loss to be finely simulated using atmospheric models. However, the computational cost prohibits the exploration of a large parameter space in operational monitoring applications. To address this, recent studies \change[AJC]{exploited}{made use of} a deep learning algorithm capable of making \add[AJC]{transmission} loss predictions almost instantaneously. However, the use of \change[AJC]{interpolated}{nudged} atmospheric models leads to an incomplete representation of the medium, and the absence of temperature as an input makes the \change[AJC]{network not adapted for}{algorithm incompatible with} long range propagation. \change[AJC]{In the current work}{In this study}, we address these limitations by using both wind and temperature fields as inputs \add[AJC]{to a neural network}, simulated up to $130$ km altitude and $4,000$ km distance. We also optimize several aspects of the \add[AJC]{neural} network architecture. We exploit convolutional and recurrent layers to capture spatially and range-dependent features embedded in realistic atmospheric models, improving the overall performance. The \add[AJC]{neural} network reaches an average error of $4$ dB compared to full parabolic equation simulations and provides epistemic and data-related uncertainty estimates. Its evaluation on the $2022$ \add[AJC]{Hunga} Tonga-Hunga Ha'apai volcanic eruption demonstrates its \change[AJC]{generalization capabilities}{prediction capability using atmospheric conditions and frequencies not included in the training}. This represents a significant step towards near real-time assessment of International Monitoring System detection thresholds of explosive sources.
\end{abstract}

\section*{Plain Language Summary}
Accurate modeling of infrasound transmission loss is essential in a wide range of applications, such as improving atmospheric data assimilation for numerical weather prediction, assessing attenuation maps of sources of interest, or estimating the \change[AJC]{detection capabilities of the International Monitoring System infrasound network}{spatial and temporal variability of the International Monitoring System infrasound network performance}. However, the high computational cost of numerical modeling solvers makes them impractical in near real-time analysis. To address this, we develop a convolutional recurrent neural network able to predict ground-level transmission losses for a propagation range of $4,000$ km and for five frequencies ranging from $0.1$ to $1.6$ Hz in $0.045$ seconds. The proposed method exploits range-dependent atmospheric specifications that combine horizontal wind speed and temperature fields, including small-scale atmospheric perturbations. In comparison with the state-of-the-art neural network \cite{qb23}, the proposed model achieves an average error of the same magnitude while extending the propagation range at a global scale and providing estimates of epistemic and data-related uncertainty. The model is evaluated on the $2022$ \add[AJC]{Hunga} Tonga-Hunga Ha'apai volcanic eruption and demonstrates its performance in generalization by providing accurate predictions at new sampling regions, new dates and new source frequencies.

%%%%%%%%%%%%%%%%%%%%%%%%%%%%%%%%%%%%%%%%%%%%%%%%%%%%%%%%%%%%%%%%%%%%%%%%%%%%%%%%%%%%%%%%%%%%%%
%%%%%%%%%%%%%%%%%%%%%%%%%%%%%%%%%%%%%%%%%%%%%%%%%%%%%%%%%%%%%%%%%%%%%%%%%%%%%%%%%%%%%%%%%%%%%%
\section{Introduction}
\label{intro}
Many high-energy atmospheric phenomena, whether natural (meteoroids, earthquakes, or volcanoes) or human-made (aircraft, chemical or nuclear explosions), generate acoustic waves at inaudible frequencies ($\le20$ Hz), called infrasound. Such waves can propagate thousands of kilometers through various atmospheric waveguides and be refracted back to the surface, allowing for detection on a global scale. \change[AJC]{For that reason, they are permanently recorded by the ground-based infrasound stations of the International Monitoring System (IMS), which monitors compliance with the Comprehensive Nuclear-Test-Ban Treaty.}{For that reason, they are continuously recorded by the ground-based infrasound stations of the International Monitoring System (IMS). In order to monitor compliance with the Comprehensive Nuclear-Test-Ban Treaty, the IMS was designed to allow the detection of any explosions with a yield of one kiloton of TNT} (\citeA{dc22}; \citeA{jm19}). \add{The International Data Center (IDC) uses automatic processes to identify such relevant events (herearfter referred to as events) among the mass of data recorded by the IMS stations, providing} an estimate of the wavefront parameters useful for source location and characterization (back-azimuth, apparent velocities, amplitudes, frequencies; \citeA{pm19}). The IMS infrasound recordings play a key role in the development of new advanced processing methods leveraging deep learning algorithms (e.g., \citeA{jb22}; \citeA{sa20}).

\smallskip

Transmission loss (TL) is the cumulative decrease in acoustic energy as waves propagate. Typically, TL is given in decibels and in terms of amplitude at range divided by the amplitude at a fixed reference distance. Accurate modeling of TL is essential to interpret the IMS infrasound \change[AJC]{array}{station} measurements and to evaluate event detection thresholds (\citeA{dg10}; \citeA{lp09}). \add[AJC]{Such evaluation represents a key step toward optimizing the design of the IMS network to effectively monitor infrasonic sources of interest (e.g., explosions) worldwide.} \change[AJC]{This can also help to better characterize the atmosphere at altitudes where other measurements are scarce, since infrasound recordings can be used to constrain various atmospheric properties such as winds and temperatures}{As infrasound recordings can provide information about various atmospheric properties, such as winds and temperatures, accurate modeling of infrasound TL can also help to better infer these properties at altitudes where direct atmospheric measurements are scarce, through inverse methods} (\citeA{ja12}; \citeA{lp05}; \citeA{ps14}; {\citeA{rv20}; \citeA{eb19}; \citeA{ja24}; \citeA{pl24}). 

\smallskip

The computational cost of existing numerical propagation modeling tools, such as finite-differences codes (\citeA{gh11}; \citeA{qb16}), spectral element methods (\citeA{qb17}; \citeA{lm21}), normal modes \cite{rwsoftware21} or parabolic equation solvers \cite{rwsoftware21}, does not allow the exploration of a wide parameter space (variations in the atmospheric state, in frequencies or source location) for near real-time TL estimation. \change[AJC]{This makes these tools inconvenient within operational settings, such as the monitoring of explosive sources by the IMS infrasound stations}{This makes these tools inconvenient for operational framework with real-time calculation requirements, such as the monitoring of explosive sources by IMS infrasound stations.} To overcome prohibitive computational time, less expressive methods as ray-tracing can be used. This approach estimates TLs at a global scale, integrating complex atmospheric data (multiple wind components, strong vertical stratification, horizontal dependencies, etc.) as well as Earth's topographical relief. However, it is limited by the presence of shadow zones (\citeA{gb39} \add[AJC]{chapter 8; }\citeA{ap19}) for which the predicted TL is underestimated. Moreover, this method is an infinite-frequency approximation which neglects diffraction-related effects that occur in finite-frequency wave propagation. \citeA{lp12} proposes an alternative approach, with empirical propagation equations relying on simplified assumptions. The model is based on multi-parametric regression fitting to the output of parabolic equation solvers to predict TL at ground-level. However, this was derived from idealized atmospheric specifications neglecting range-dependent variations and focusing only on wind speed and temperature values at ground-level and at 50 km altitude above a station. As a result, \change[AJC]{the model struggles to predict TL in cases when the atmospheric waveguiding varies with distance. Moreover, errors as large as $70$ dB can be reached in the absence of stratospheric waveguides or at high frequencies ($\ge1$ Hz).}{the model struggles to predict TL when the structure of the atmospheric waveguides varies along the propagation path, in the absence of stratospheric guides, or at high frequencies ($\ge1$ Hz)}.

\smallskip

Machine learning techniques are currently explored in the field of acoustic wave propagation modeling in the atmosphere. \citeA{qb23} enhances the work of \citeA{lp12} by developing a deep learning algorithm predicting infrasound TL from realistic range-dependent atmospheric fields simulated \change[AJC]{on}{at} a regional scale. The method exploits a convolutional neural network \cite{yl15} to estimate ground-level TL at around $5$ dB error compared to parabolic equation simulations. A key aspect is the negligible computation times at the inference stage (around $0.05$ s per prediction) compared to PE solvers. The TL is estimated for a given frequency between $0.1$ and $3.2$ Hz. While these results are promising, the propagation range of $1,000$ km is a limitation when performing global-scale investigation of TL, e.g., when studying long-range propagation of microbarom \cite{ev15} or when predicting the event detection capability of the IMS, where the average distance between infrasound stations is around $2,000$ km. Moreover, the input data used to train the model are built using an interpolation of various atmospheric specifications with different resolution, which can lead to an incomplete resolution of the atmospheric variability. A simplification is finally made by considering only wind speed values as inputs of the neural network.

\citeA{cp20} introduced a physics-aware neural network to model acoustic wave propagation through the atmospheric boundary layer ($\le2$ km altitude) and recover underlying physical parameters. The propagation range is kept within $1$ km from the source and only audible frequencies are considered ($50$ and $150$ Hz). The proposed model is a Physics-Informed Neural Network. It is a data-driven algorithm accounting for physical constrains by integrating an additional regularization term in its loss function. This approach \change[AJC]{approximates well}{provides a good approximation of} the TL fields at a local scale but fails to recover the underlying physical parameters not provided as inputs (friction velocity, surface heat flux, etc.). Another study developed a fully-connected neural network to model two-dimensional acoustic TL from a set of predefined inputs describing a turbulent atmosphere \cite{ch21}. Errors of about $7$ dB were obtained over $10$ km distance, but the atmospheric modeling is reduced to $13$ classes, which does not allow for a detailed characterization of the medium. Finally, \citeA{fl24} demonstrated the capabilities of Fourier neural operator (\citeA{lz20}) algorithms to model seismic wave propagation. However, the considered propagation medium has much less variability than the atmosphere.

\smallskip

\add[AJC]{In this study,} we enhance the work of \citeA{lp12} and address the limitations of \citeA{qb23} by exploiting realistic range-dependent wind speed and temperature fields simulated at a global scale \add[AJC]{(up to $4,000$ km distance)} and by optimizing the deep learning architecture. The key optimization relies on the use of recurrent layers in addition to convolutional layers. The convolutional layers capture the spatially local features embedded in the input atmospheric specifications, while the recurrent ones catch the range dependencies; as infrasound waves \change[AJC]{go}{travel} only in the "forward direction" so that the attenuation at a range $d$ cannot depend on the atmospheric state beyond. The proposed neural network is trained on a large set of \add[AJC]{$4,000$ km-long} parabolic equation simulations. Predictive uncertainties are quantified by considering the uncertainty related to the model architecture and to the input data. We retrospectively apply our model on the atmospheric conditions of January $15$, $2022$ to emulate a near real-time TL map \change[AJC]{for infrasound generated by the major volcanic eruption of the Tonga-Hunga Ha'apai (Tonga).}{around the Hunga Tonga-Hunga Ha'apai (now referred to as Tonga) volcano, whose major eruption generated infrasound waves that propagated over several thousand kilometers}  \cite{jv22}. The network is even evaluated on new source frequencies, representing a concept validation and a first step towards a near real-time evaluation of the detection threshold of the IMS infrasound network \add[AJC]{at a global scale}. The current paper is organized as follows. In Section~\ref{section_2}, the data used to train the neural network are presented. Section~\ref{section_3} details the architecture of the Convolutional Recurrent Neural Network. Section~\ref{section_4} evaluates the model performances on testing and generalization data. Section~\ref{section_5} quantifies the epistemic and the data uncertainties. Finally, we discuss and summarize our findings in Section~\ref{section_6} and Section~\ref{section_7}, which includes the Tonga case study.

%%%%%%%%%%%%%%%%%%%%%%%%%%%%%%%%%%%%%%%%%%%%%%%%%%%%%%%%%%%%%%%%%%%%%%%%%%%%%%%%%%%%%%%%%%%%%%
%%%%%%%%%%%%%%%%%%%%%%%%%%%%%%%%%%%%%%%%%%%%%%%%%%%%%%%%%%%%%%%%%%%%%%%%%%%%%%%%%%%%%%%%%%%%%%
\section{Data for transmission loss estimation}
\label{section_2}

\subsection{Inputs: realistic range-dependent atmospheric slices}
\label{input_atmospheric_slices}
Infrasound propagation is sensitive to the atmospheric medium, in particular to horizontal and vertical gradients of sound speed, which is directly related to temperature and wind \cite{gh09}. Waves traveling through the atmosphere can be refracted down to the ground by positive gradients or upwards by negative ones. Hence, infrasound can propagate through different waveguides, depending on the wind and the temperature in the troposphere ($0$--$12$ km altitude), stratosphere ($12$--$60$ km altitude) and mesosphere ($60$--$90$ km altitude). Above $90$ km altitude, in the thermosphere, the temperature greatly increases, leading to a permanent thermospheric waveguide. Within the geometric ray-tracing approximation, the effective sound speed ratio $c_\text{ratio}$ indicates the presence or the absence of such guides:   
\begin{equation}
    c_\text{ratio}(z)=\frac{c_\text{eff}(z)}{c_\text{eff}(z=0)} \quad ; \quad c_\text{eff}(z)=u_0(z) + c(z),
    \label{eq:1}
\end{equation} 

where $u_0(z)$ is the horizontal component of the wind speed in the direction of propagation and $c(z)=\sqrt{\gamma R T(z)}$ is the adiabatic sound speed with $\gamma$ the adiabatic index, $R$ the specific gas constant for air and $T$ the absolute temperature. For sources at ground-level and a given altitude $z$, the condition $c_\text{ratio}(z)\ge 1$ indicates the presence of a guide refracting the wave back to the ground, turning at the altitude $z$. This approximation holds only for high frequencies, flows with small Mach numbers and shallow propagation angles. It does not take into account the influence of cross-winds or diffraction. However, as shown in \citeA{ja12}, infrasound TL is not very sensitive to such winds. 

\smallskip

The atmospheric model input data used to train the neural network are realistic slices $A_{z,d}$ of $c_\text{ratio}$. The slices are vertical-horizontal planes covering the altitude $z \in [0, 130]$ km and the distance from source $d \in [0, 4000]$ km (see Figure~\ref{fig:1}). The temperature and horizontal wind speed values used to calculate the $c_\text{ratio}$ are extracted from the Whole Atmosphere Community Climate Model (WACCM) forecast products (Atmospheric Chemistry Observations \& Modeling, National Center for Atmospheric Research, University Corporation for Atmospheric Research; \citeA{ag19}). We use the sixth version of this product, providing a horizontal resolution of ($100 \times 140$) km and a model top above $130$ km, distributed across $88$ vertical levels. In contrast, \citeA{qb23} used the ERA$5$ high-resolution re-analysis product from the European Centre for Medium-Range Weather Forecasts to extract wind speed values up to the mesosphere. This model possesses $137$ vertical levels but reaches a maximum altitude of $80$ km \cite{hh05}. To represent the atmosphere beyond this limit, \citeA{qb23} used two climatological specifications. The specification HWM-$14$ allows retrieving the horizontal wind speed fields \cite{dd15} and the model NRLMSIS-$00$, the temperature values \cite{jp02}. ERA$5$ and the climatological models were connected together using a cubic interpolation between $75$ and $85$ km altitude. In comparison to WACCM, this hybrid approach does not capture the interannual variability in the upper layers of the atmosphere. This can lead to \change[AJC]{less well}{poorly} modeled wave propagation in the absence of stratospheric guides. 

\begin{figure}               
    \includegraphics[width=3.8in]{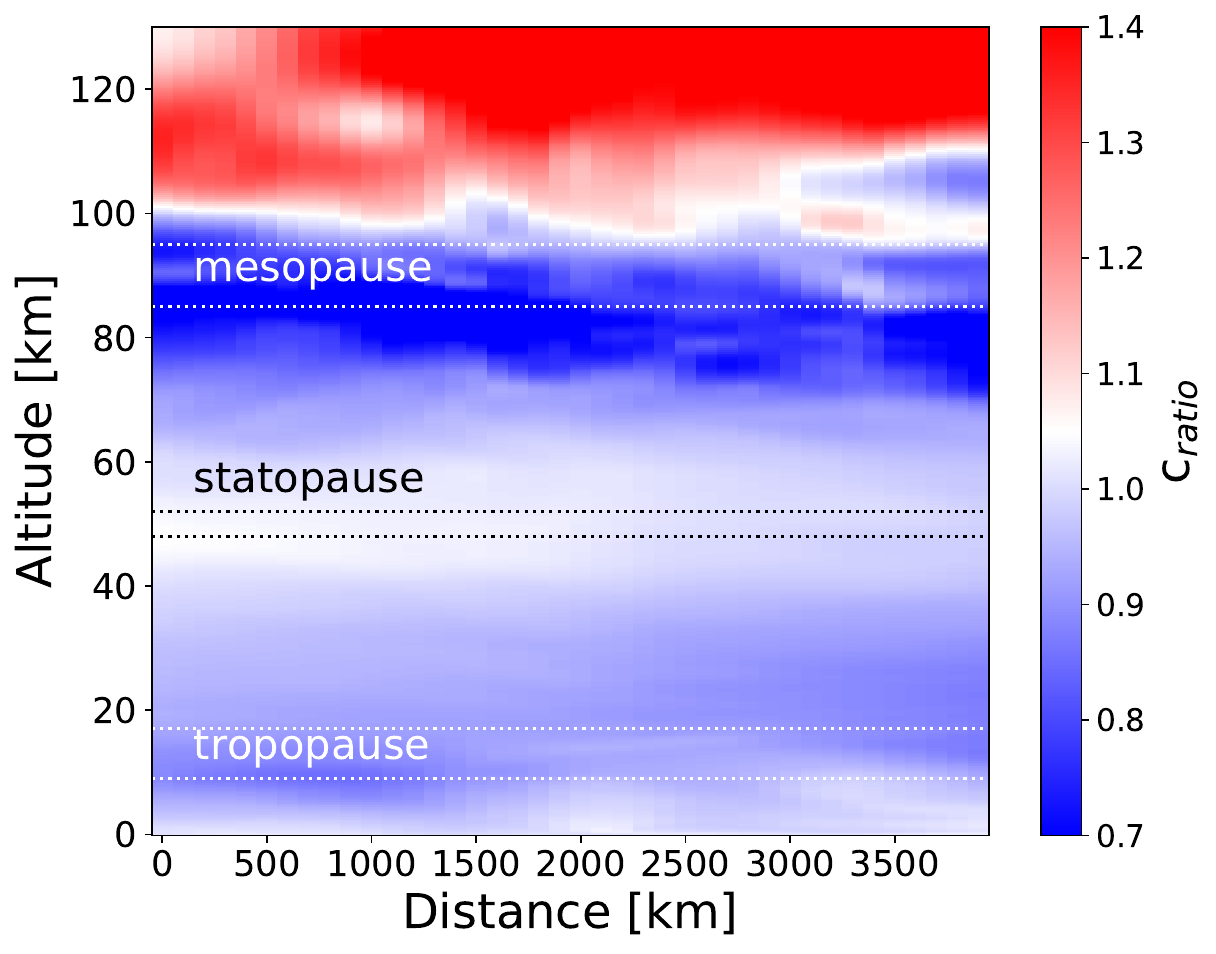}
    \caption{Two-dimensional $c_\text{ratio}$ field constructed using temperatures, zonal and meridional wind speeds extracted from the WACCM specification (no GW perturbations included). This example highlights the waveguide due to the strong increase of the temperature above the mesopause as well as a weaker guide in the stratosphere ($c_\text{ratio}\ge1$ in red and white, respectively).} 
    \label{fig:1} 
\end{figure}

\smallskip

\label{gardner}
Because of its coarse spatial resolution, WACCM cannot resolve the full gravity wave (GW) spectrum, and the effect of these waves must be parameterized as it is the case in most state-of-the-art numerical weather prediction systems (\citeA{ag19}; \citeA{ma21}). However, such parameterizations represent the deposition of GW energy in the larger scale flow without accounting for the vertical propagation of the waves \cite{pl20}. Notably, the perturbations of the wind and temperature fields along the GW propagation paths are not accounted for when GW are not resolved by the dynamical core. As these oscillations can have a major impact on infrasound propagation paths \cite{ic19}, \change[AJC]{it is important}{attention should be paid} to account for their effect in propagation simulations, and in particular across the IMS infrasound network \cite{cl24}. In the current work, we opt for an idealized albeit already used method in infrasound studies (\citeA{dg10}; \citeA{lp12}; \citeA{qb23}) which involve adding perturbations with realistic amplitudes obtained from a vertical GW spectrum model \cite{cg93}. \add[AJC]{Figure} \ref{fig:2} \add{shows an example of a vertical $c_\text{ratio}$ profile disturbed by GW perturbations (panel a)) as well as a two-dimensional range-dependent GW perturbation field (panel b)), which can be superimposed on the $c_\text{ratio}$ slices. See} \ref{appendix_1} \add{for additional details on the process developed to obtain such perturbations. This approach assumes a continuum (background) of sources and does not account for the filtering of GWs by the large-scale flow upon vertical propagation. However, it provides the needed small scales of realistic amplitudes to feed the training database with relevant atmospheric features otherwise lacking from the WACCM atmospheric model used in the current study. Moreover, the stochastic generation of GW through the randomly chosen phases (see} \ref{appendix_1}\add{) leads to an ensemble approach. In doing so, we define an uncertainty for the atmospheric field, which is used in Section} \ref{section_5} \add{where the sensitivity of the model to input data is investigated.}

\begin{figure}  
    \includegraphics[width=3.8in]{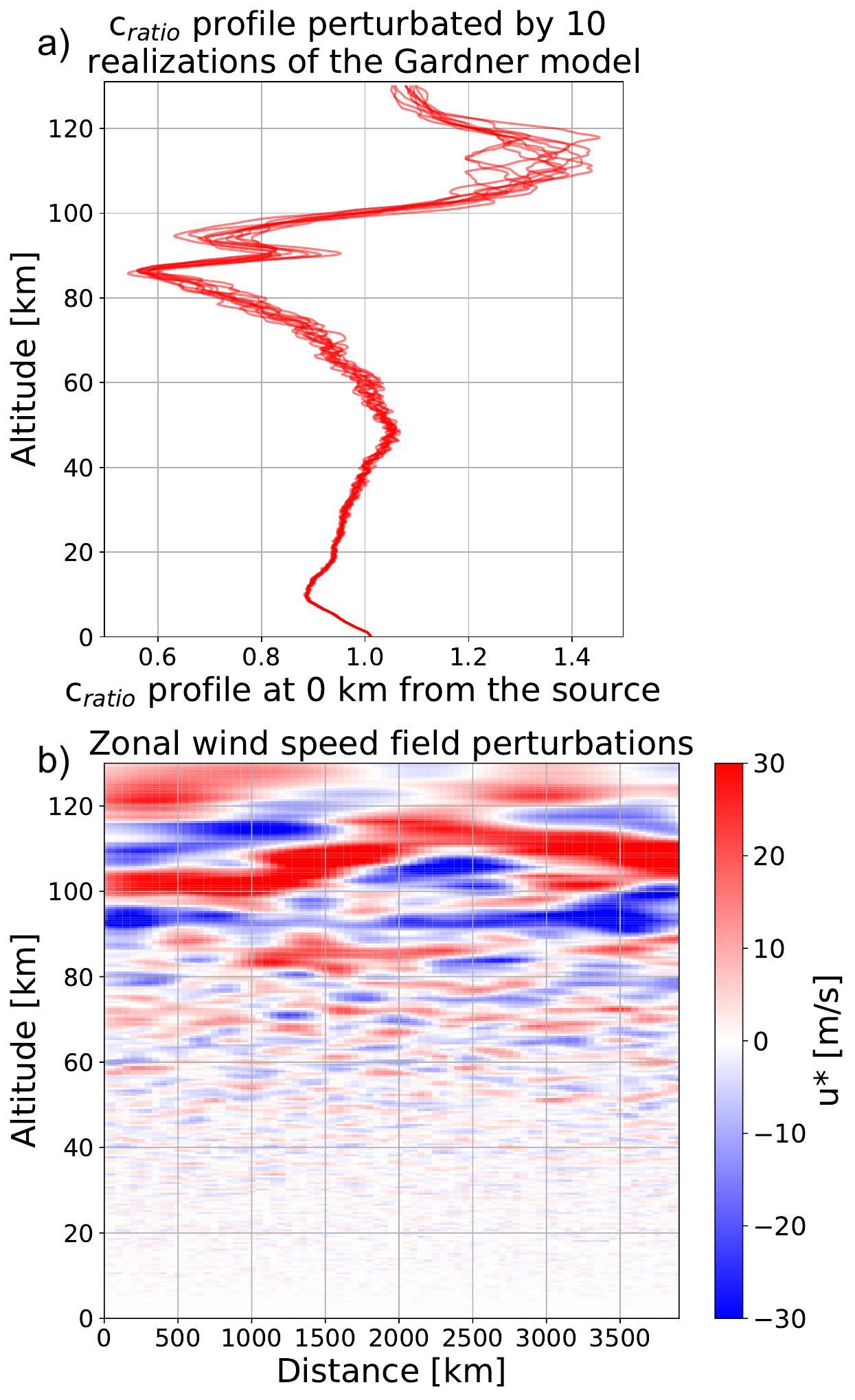}
    \caption{a) A vertical $c_\text{ratio}$ profile perturbed by an ensemble of ten realizations of the GW model. b) Two-dimensional zonal wind disturbances field induced by the GW. Ten realizations of such perturbation fields are added to each slice of $c_\text{ratio}$.} 
    \label{fig:2} 
\end{figure}

\smallskip

Aiming to represent a large quantity of atmospheric conditions, we sample the Earth with a set $162$ points arranged on a regular grid of $20$°-resolution on January $15$, $2021$. From each point, we collect atmospheric slices $A_{z,d}$ of $c_\text{ratio}$ along eight directions: north, north-east, east, south-east, south, south-west, west and north-west. Each slice is perturbed by ten two-dimensional GW fields and is projected along two azimuths, $90$° and $270$°. We therefore obtain $K=162 \times 8 \times 10 \times 2=25,920$ atmospheric slices $A_{z,d}$. \add[AJC]{The choice of spatial resolution results from a trade-off between achieving global sampling and maintaining a reasonable computational cost for building the database. Moreover, propagating along eight different directions and projecting onto two azimuths enables the multiplication of atmospheric conditions in a second step, at minimal additional computational cost.}

%%%%%%%%%%%%%%%%%%%%%%%%%%%%%%%%%%%%%%%%%%%%%%%%%%%%%%%%%%%%%%%%%%%%%%%%%%%%%%%%%%%%%%%%%%%%%%
\subsection{Outputs: infrasound transmission losses}
\label{output_transmission_losses}
The TL expresses the amplitude and phase variation as a wave propagates through the layered structure of the atmosphere. Along its path, a part of the acoustic energy loss is due to intrinsic wave attenuation mechanisms in the atmosphere, combined with the geometric spread. The TLs to be estimated are simulated using the numerical solver \textit{ePape} from the NCPAprop package developed at the National Center for Physical Acoustics \cite{rwsoftware21}. \textit{ePape} simulates, for a single frequency and in a vertical-horizontal plane, the amplitude of the long-range pressure produced by a unit point source, in relation to the level at a reference distance of $1$ km from that source. The TL is obtained by calculating the modulus of this simulated pressure field and is commonly expressed in decibels:
\begin{equation}
    \text{TL}=20 \times \text{log}_{10}\bigg(\frac{P}{P_0}\bigg),
    \label{eq:2}
\end{equation} 

where $P$ is the pressure at a given distance and $P_0$ the pressure at the reference distance of 1 km. 

\smallskip

\textit{ePape} uses the Sutherland-Bass atmospheric absorption coefficients \cite{ls04} and the parabolic equation method (PE) to simulate the pressure field. The PE method is well adapted to simulate wave propagation in a range-dependent medium -- such as the one modeled by the atmospheric slices $A_{z,d}$ -- by accounting for diffraction effects and scattering by acoustic impedance variations in the atmospheric model \cite{rw19}. The PE method assumes the atmospheric medium as a locally stratified domain and considers only signals with a frequency content large compared to the Brunt-Väisälä frequency ($0.05$ Hz). Inputs to the \textit{ePape} solver are atmospheric states represented by the zonal, meridional and vertical winds in $\text{m.s}^{-1}$, the temperature in Kelvin, the density in $\text{g.cm}^{-2}$ and the pressure in mbar. The along-path winds are calculated by projecting the zonal and meridional components along the wave propagation path from the source to the receiver. The influence of the along-path and vertical winds on the propagation is modeled using the effective sound speed approximation described in Section~\ref{input_atmospheric_slices}. 

\smallskip

Each of the $25,920$ atmospheric slices $A_{z,d}$ is associated to five ground-level TLs $\ell_f$ simulated at frequencies $f=0.1, 0.2, 0.4, 0.8$ and $1.6$ Hz. The maximum propagation range is fixed at $4,000$ km from the source. \add[AJC]{This enables the study of events on a global scale (such as the Tonga volcano eruption), as well as the evaluation of the detection capabilities of the IMS network, whose infrasound stations are spaced approximately $2,000$ km apart.} The expensiveness of the PE method only appears during the database creation stage. The proposed neural network will predict ground-level TL from new atmospheric scenarios almost instantaneously.

%%%%%%%%%%%%%%%%%%%%%%%%%%%%%%%%%%%%%%%%%%%%%%%%%%%%%%%%%%%%%%%%%%%%%%%%%%%%%%%%%%%%%%%%%%%%%%
%%%%%%%%%%%%%%%%%%%%%%%%%%%%%%%%%%%%%%%%%%%%%%%%%%%%%%%%%%%%%%%%%%%%%%%%%%%%%%%%%%%%%%%%%%%%%%
\section{A Convolutional Recurrent Neural Network-based solution for TL estimation}
\label{section_3}

\subsection{Architecture of the surrogate model}
To accurately and rapidly estimate ground-level TL, we develop a supervised neural network designed to emulate the output of the numerical solver \textit{ePape}. We denote $F_\theta(A_{z,d},f)$ this network. Supervised neural networks are powerful learning systems capable of approximating non-linear functions by optimizing a set of parameters $\theta$ (weights and biases) to minimize a loss function. They map input data, such as $(A_{z,d},f)$, to output data, such as $\ell_f$, by learning hierarchical representations that capture the complex relationships between inputs and outputs.

\begin{figure}
    \includegraphics[width=3.75in]{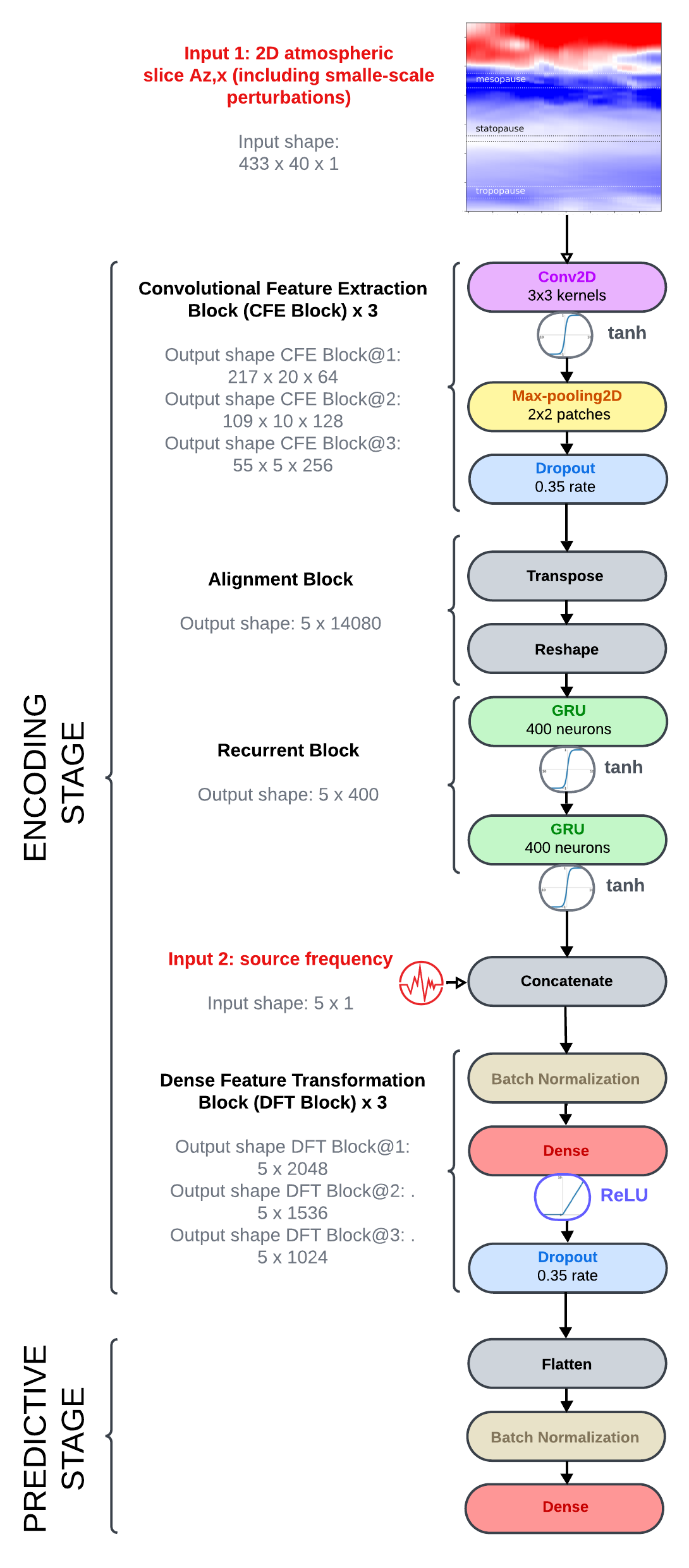}
    \caption{Architecture of $F_\theta(A_{z,d},f)$. The atmospheric model $A_{z,d}$ including GW perturbations and the source frequency $f$ are first encoded by three CFE Blocks, an Alignment Block, a Recurrent Block and three DFT Blocks. Then, the ground-level TL is predicted over $800$ points ($4,000$ km).}
    \label{fig:3}
\end{figure}

\smallskip

The proposed deep learning architecture relies on Convolutional Neural Networks and Recurrent Neural Networks (CNNs and RNNs; \citeA{yl15}). CNNs are well adapted for extracting spatial patterns in multidimensional data. They detect and encode local spatial correlations by applying convolutional filters that extract features. These features are progressively transformed and compressed into a latent space of reduced spatial dimensions through convolution and pooling operations. This latent space captures increasingly abstract and informative representations of the input, making CNNs particularly effective for identifying spatial dependencies in the two-dimensional arrays $A_{z,d}$. RNNs process sequential data by maintaining a state vector, which serves as a memory of past information in the sequence. At each time step, the step vector is updated by combining the current input with the information from previous steps, allowing the network to model temporal or range dependencies. However, traditional RNNs struggle with long-range dependencies due to vanishing or exploding gradients. To address this limitation, Gate Recurrent Units (GRUs; \citeA{kc14}) are applied in the current work. GRUs employ gating mechanisms to control the flow of information, enabling the network to capture both short- and long-range dependencies efficiently. In the context of this study, GRUs are well-adapted for processing range-dependent atmospheric slices, where sequential correlations between adjacent $c_\text{ratio}$ profiles play a critical role in estimating TLs. 

\smallskip

To detect and encode relevant features in the atmospheric arrays $A_{z,d}$, $F_\theta(A_{z,d},f)$ stacks three "Convolutional Feature Extraction Blocks" (CFE Blocks), an "Alignment Block", a "Recurrent Block", and three "Dense Feature Transformation Blocks" (DFT Blocks). Each CFE Block contains a two-dimensional convolutional layer, a pooling layer and a dropout layer. The convolutional layers (in purple in Figure \ref{fig:3}) have an increasing number of filters across the CFE Blocks ($64$, $128$, and $256$, respectively). These filters progressively extract more complex and abstract features from the input data. We incorporate non-linearity in $F_\theta(A_{z,d},f)$ by using the smooth and differentiable hyperbolic tangent activation function. Hyperbolic tangent transforms output values between $-1$ and $1$, which limits the risk of gradient explosion. The max-pooling layers (in yellow in Figure \ref{fig:3}) perform dimensionality reduction on the convolved arrays $A_{z,d}$. They limit the number of parameters to be learned and create invariance to small shifts and distortions. Finally, dropout layers (in blue in Figure \ref{fig:3}) randomly drop $35$\% of the connections between neurons in each forward pass. They introduce a regularization that limits overfitting by making the network less dependent on specific neurons. At the end of the three CFE Blocks, the atmospheric slices $A_{z,d}$ are encoded into latent structures of dimension $(\text{height} \times \text{width} \times \text{depth})=(55 \times 5 \times 256)$, the \textit{height} being the altitude and the \textit{width} the distance from the source. 

\smallskip

Each latent structure is then passed to the Alignment Block to be transposed and reshaped into $(5 \times (55 \times 256))$ arrays. This allows them to be exploited by two GRU layers (in green in Figure \ref{fig:3}) which require inputs in the form of (\textit{time steps} $\times$ \textit{features}). The recurrent layers improve the performance of the architecture by enabling the range dependencies that exists in the encoded range-dependent atmospheric slices to be captured. The source frequency $f$ of the TL to be predicted is added at the end of the Recurrent Block.

\smallskip

The last transformations applied on the encoded atmospheric slices and the source frequency are performed by three DFT Blocks. Each DFT Block contains a batch normalization layer, a fully-connected layer, and a dropout layer. Batch normalization (in brown in Figure \ref{fig:3}) keeps the data standard deviation close to $1$, which stabilizes the training and increases the robustness. The fully-connected layers (in red in Figure \ref{fig:3}) have a decreasing width across the DFT Blocks, containing respectively $2048$, $1536$ and $1024$ neurons. As normalization operations are performed before each fully-connected layer, we use the Rectified Linear Unit function as the activation function (ReLU; \citeA{xg11}) without risking gradient explosion even if it does not hold the data distribution between $-1$ and $1$ as the hyperbolic tangent function does.

\smallskip

After these encoding stages, $F_\theta(A_{z,d},f)$ flattens and normalizes the data to pass them to a $800$-neurons fully-connected layer with the linear activation function to estimate the ground-level TL $\ell_f$ linearly interpolated over $800$ points.

%%%%%%%%%%%%%%%%%%%%%%%%%%%%%%%%%%%%%%%%%%%%%%%%%%%%%%%%%%%%%%%%%%%%%%%%%%%%%%%%%%%%%%%%%%%%%%
\subsection{Learning procedure}
\label{learning}
To optimize memory, we interpolate all atmospheric slices $A_{z,d}$ and ground-level TL. The atmospheric slices are interpolated to a regular grid $(\text{height} \times \text{width} \times \text{depth})=(433 \times 40 \times 1)$. The \textit{height} corresponds to the altitude, with $z \in [0, 129.9]$, and the \textit{width} corresponds to the distance from the source, with $d \in [0, 3900]$; keeping only a vertical $c_\text{ratio}$ profile every $100$ km. The resulting $0.3$ km step in altitude is sufficient to preserve the effects induced on the atmospheric slices by the small-scale GW perturbations. Indeed, as stated in \remove[AJC]{Section 1}\ref{appendix_1}, the GW vertical spectra upon which the perturbations are built peak at a critical wavenumber $m^*$ corresponding to a few kilometers in wavelength. The exponential decrease of energy at smaller and larger wavenumbers, respectively, ensures that the relevant part of the GW spectrum will be well accounted for at each altitude with a $300$ m vertical step. The ground-level TL is interpolated with a regular step of $5$ km, leading to $\ell_f$ vectors of $800$ points. This allows unifying  the TL calculated using the \textit{ePape} solver, which has different spatial resolution according to the source frequency $f$. 

\smallskip

Both the atmospheric slices $A_{z,d}$ and ground-level TL $\ell_f$ are standardized before feeding $F_\theta(A_{z,d},f)$. The standardization involves removing the mean and scaling to unit variance. This pre-processing step allows us to speed up learning and minimize the risks of explosion/disappearance of the gradient when dealing with an important number of parameters ($F_\theta(A_{z,d},f)$ contains approximately twenty-seven million parameters). \label{standardization}

\smallskip

\label{training_validation_datasets}
To train $F_\theta(A_{z,d},f)$, we set aside some of the atmospheric slices $A_{z,d}$ and associated PE simulations $\ell_f$ as training and validation datasets. Five GW realization fields out of ten are superimposed on each slice $A_{z,d}$, keeping half of the small-scale variations field unseen during the learning to evaluate the model on them later. Within the same objective, we remove $12$ sampling points (all eight directions of propagation associated) among the initial $162$ ones. $70$\% and $20$\% of the remaining data are randomly selected to form the training and the validation datasets (see red and green lines in Figure \ref{fig:4}). As a result, the training dataset contains $K_{\text{train}}=0.7 \times [(162-12) \times 8 \times 5 \times 2 \times 5]=42,000$ samples. We perform a cross-validation with ten independent selections of the training and validation samples. This allows training the surrogate model ten times (each training being referred as a "training run"), and keeping the model with the best hyper-parameters. 

\begin{figure}
    \includegraphics[width=6.1in]{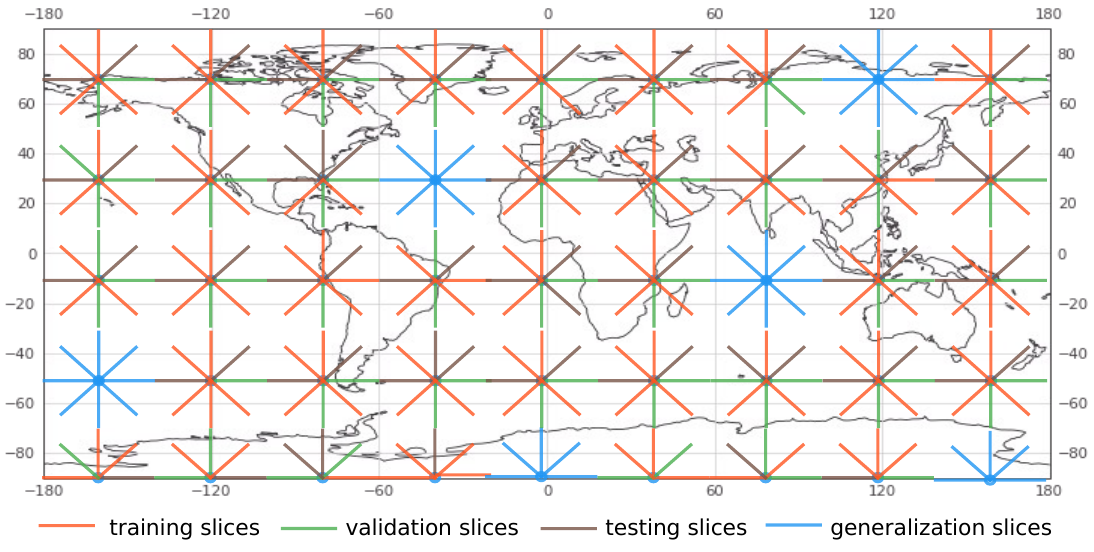}
    \caption{Illustration of the division of the atmospheric slices $A_{z,d}$ into training (red), validation (green), test (brown), and generalization (blue) samples.} 
    \label{fig:4} 
\end{figure}

\smallskip

Following the work of \citeA{qb23}, the difference between the predictions $\hat{\ell}_f$ and the PE simulations $\ell_f$ is quantified by the Root Mean Square Error (RMSE) loss function. This metric, widely used in regression problems, is defined as:
\begin{equation}
    \text{RMSE}=\sqrt{\text{MSE}}=\sqrt{\frac{1}{K} \sum_{k=1}^{K}|\ell_{f,k}-\hat{\ell_f}_{k}|^2},
    \label{eq:4}
\end{equation} 

where $K$ is the number of samples in the dataset. Since the set of atmospheric slices $A_{z,d}$ and the set of expected outputs $\ell_f$ are standardized, the RMSE is dimensionless. The training losses of the ten training runs of $F_\theta(A_{z,d},f)$ decrease similarly during the learning, converging to an RMSE of $0.12$. The lowest RMSE value of $0.119$ is reached after $76$ iterations by the first training run (see red line in Figure \ref{fig:5}). Validation loss gives a qualitative understanding of the generalization capabilities of the model. The only training run reaching a validation RMSE minimum below $0.1$ is the eighth one, with a best RMSE value equals to $0.098$. Nevertheless, for the rest of this contribution, we focus on the first training run which provides, on average, the best training and validation metrics. 

\begin{figure}
    \includegraphics[width=3.85in]{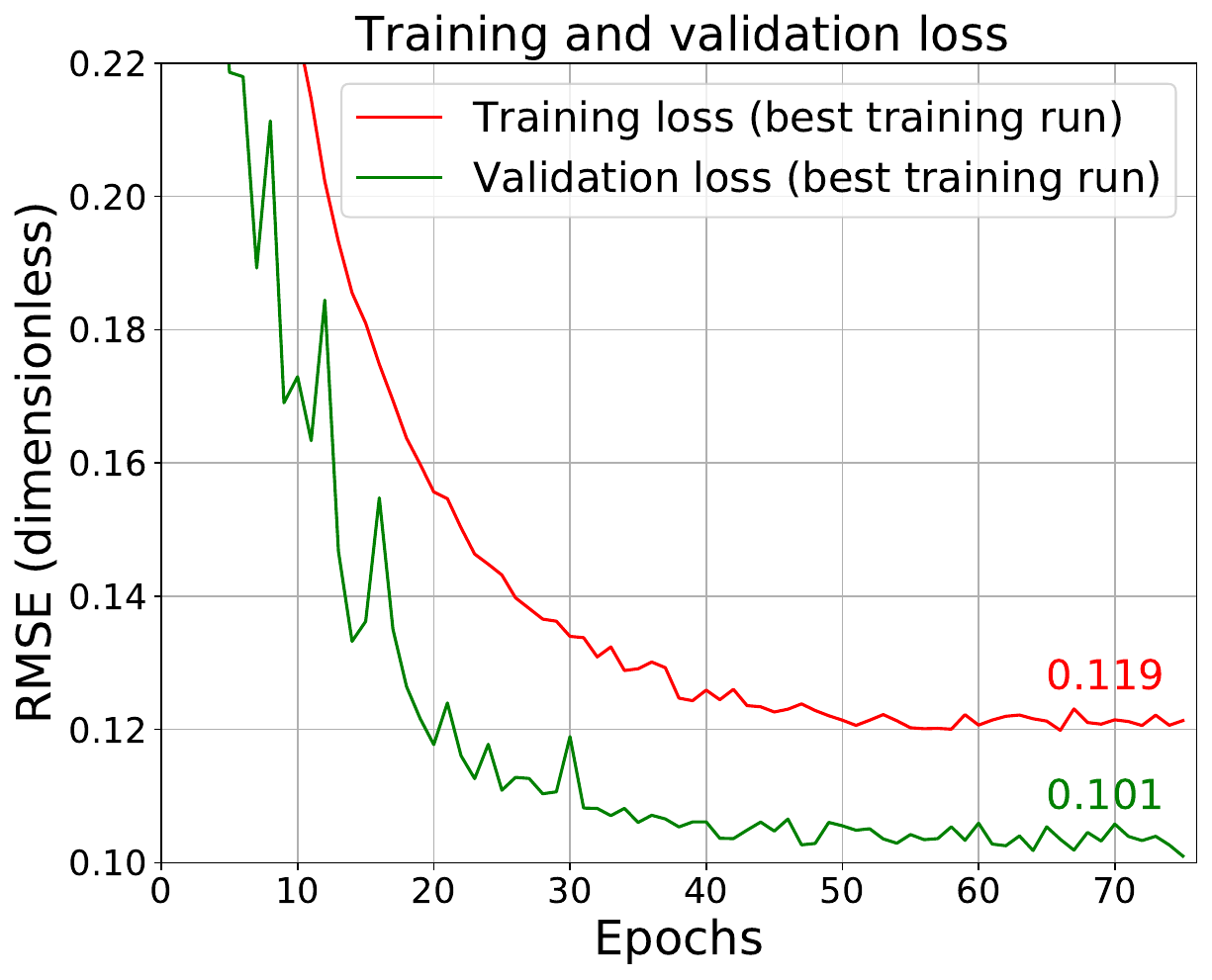}    
    \caption{Evolution of the training loss (in red) and the validation loss (in green) of $F_\theta(A_{z,d},f)$ during the best training run.} 
    \label{fig:5} 
\end{figure} 

\smallskip

The neural network $F_\theta(A_{z,d},f)$ is implemented in Python using the TensorFlow library \cite{masoftware15}. Training is conducted over a maximum of $150$ epochs, with an early stopping if the loss function on validation data remains stable beyond $20$ iterations. To improve the convergence of the loss function, the parameters $\theta$ are initialized in the first convolutional layer using the Glorot initializer \cite{xg10} and the input data $(A_{z,d},f)$ are divided into batches of size $32$. The Adam optimizer \cite{dk14} is used with an initial learning rate $\eta=1 \times 10^{-4}$, which is reduced by a factor of $10$ after the tenth epoch if no improvement in validation loss is observed. The training was performed on a high-performance computing cluster equipped with Nvidia A$100$ GPUs ($40$ GB memory each). On average, \change[AJC]{train}{training} the model takes approximately $23$ minutes for each of the ten independent training runs. After training, $F_\theta(A_{z,d},f)$ achieves near-instantaneous inference. It predicts a $4,000$ km-long ground-level TL in $0.045$ seconds regardless \add[AJC]{of} the source frequency $f$ (Dell Inc. Intel(R) Core(TM) i$9$-$13900$ $48$ CPUs $77.8$ GB RAM on RedHat $9.5$). By gathering multiple samples in the same batch, it can reduce this time to approximately $0.08$ seconds for $100$ simultaneous predictions. The estimated computation time saving compared to the PE method is three to four (for $f \ge 3$ Hz) orders of magnitude for global-scale applications requiring several thousand simulations at various frequencies (as in the assessment of detection capabilities of the IMS network).

%%%%%%%%%%%%%%%%%%%%%%%%%%%%%%%%%%%%%%%%%%%%%%%%%%%%%%%%%%%%%%%%%%%%%%%%%%%%%%%%%%%%%%%%%%%%%%
%%%%%%%%%%%%%%%%%%%%%%%%%%%%%%%%%%%%%%%%%%%%%%%%%%%%%%%%%%%%%%%%%%%%%%%%%%%%%%%%%%%%%%%%%%%%%%
\section{Experimental evaluations}
\label{section_4}

\subsection{Performances on the testing dataset}
\label{results_test}
In this section, we evaluate the neural network $F_\theta(A_{z,d},f)$ on the testing data. We investigate the impacts of the initial atmospheric conditions and the wave frequency. To quantify the average performance, two metrics are introduced (the Relative Absolute Error and the Root Mean Squared Error both averaged along the propagation path). 

\smallskip

We evaluate $F_\theta(A_{z,d},f)$ on the $10$\% remaining data limited to five GW perturbation fields per atmospheric slice (see brown lines in Figure \ref{fig:4}). At this stage, $12$ sampling points among the $162$ \remove[AJC]{ones} are kept aside to use them later for the generalization. As a result, the testing dataset contains $K_{\text{test}}=0.1 \times [(162-12) \times 8 \times 5 \times 2 \times 5]=6,000$ samples. 

\smallskip

\change[AJC]{Each predicted and simulated TL is de-standardized back into decibels.}{Each predicted and simulated TL is de-standardized by reversing the process described in Section}~\ref{standardization}\add{. This allows us to re-express the TLs in their original unit (decibels) after being standardized during the training process, and to ensure the physical interpretability of the results.} Figure \ref{fig:6} shows three examples of TL for various frequencies and atmospheric conditions. Panel a) corresponds to a downwind scenario (horizontal average of $c_\text{ratio}$ maxima between $30$ and $60$ km altitude greater than one), which favors long-range propagation of infrasound. Panel b) illustrates a typical upwind scenario with an horizontal average of $c_\text{ratio}$ maxima between $30$ and $60$ km altitude lower than one. In both cases, $F_\theta(A_{z,d},f)$ approximates well the TL as a function of distance as well as the behavior far from the source. The model also captures quite well the first shadow zone visible in panel a), when the propagation conditions are favorable in the stratosphere. One advantage of $F_\theta(A_{z,d},f)$ is its ability to represent rapid changes in propagation regimes across the distance, leading to strong variations in the TL to be predicted along the propagation path. This ability is missing in the semi-analytical attenuation model developed by \citeA{lp12}. However, panel c) nuances this observation, with an example of prediction misfitting the PE simulation after $2,500$ km of propagation. Finally, we observe in all cases that $F_\theta(A_{z,d},f)$ does not reproduce all small-scale spatial variations present in the PE simulations. We explain this "low pass filter" property by the well-known difficulty of deep learning algorithms in predicting high-frequency features (referred to as "spectral bias"; \citeA{na19}) and by the use of convolutional layers which tend to smooth the local patterns. \label{spectral bias}

\begin{figure}
    \includegraphics[width=3.15in]{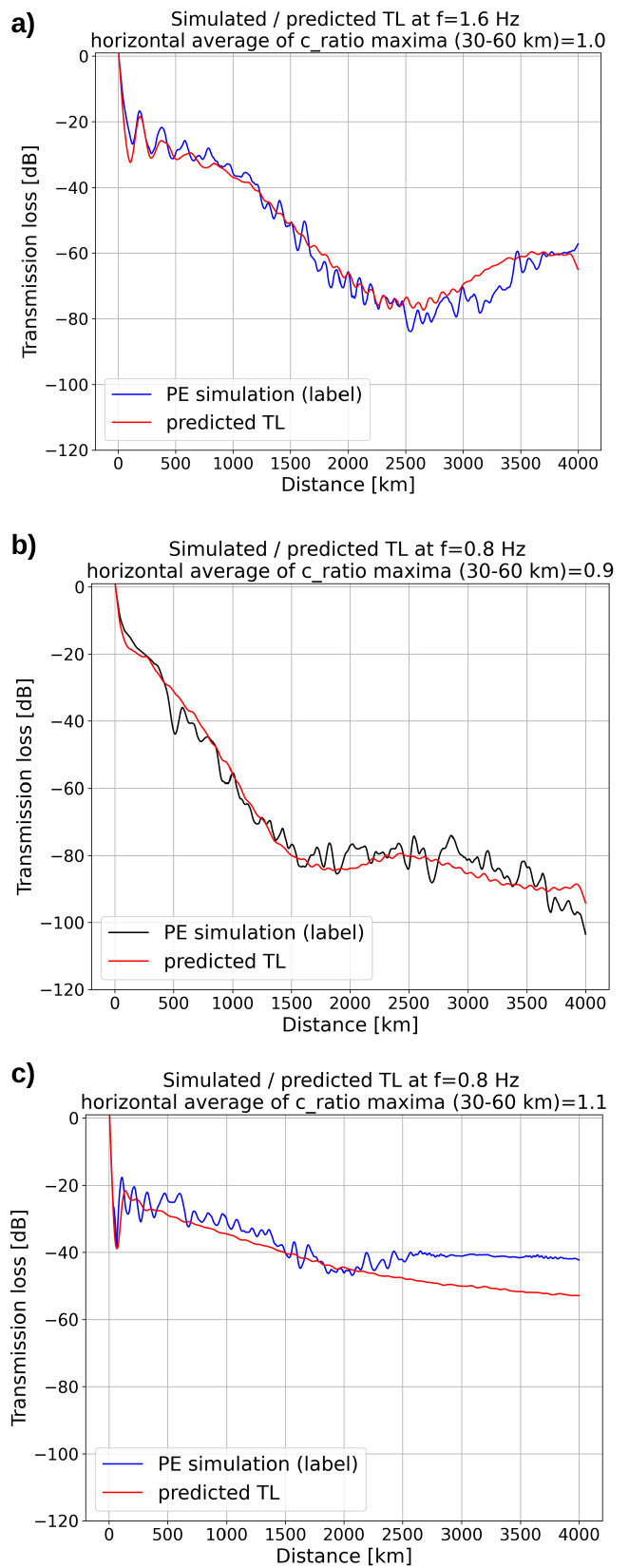}
    \caption{Examples of ground-level TLs predicted by $F_\theta(A_{z,d},f)$ and compared to full PE simulations. The maximum of $c_\text{ratio}$ between $30$ and $60$ km altitude is averaged over $4,000$ km. a) Results at $1.6$ Hz with $c_\text{ratio}=1$ (downwind scenario). b) Results at $0.8$ Hz with $c_\text{ratio}=0.9$ (upwind scenario). c) Poorer result at $0.8$ Hz.}
    \label{fig:6} 
\end{figure}

\smallskip

A broader evaluation of the testing dataset has been carried out to confirm the observation derived from Figure \ref{fig:6}. Figure \ref{fig:7} compares the $6,000$ predictions made by $F_\theta(A_{z,d},f)$ to the PE simulations and quantifies the predictive error. In the three panels of the figure, scenarios are horizontally arranged, sorted on the y-axis by the horizontal average of $c_\text{ratio}$ maxima between $30$ and $60$ km. This representation allows investigating the ability of $F_\theta(A_{z,d},f)$ to capture stratospheric guides influence on wave propagation \cite{gh09}. The comparison between the PE simulations (panel a)) and the predictions (panel b)) confirms the capacity of the model to recover the overall TL, while smoothing out fine-scale spatial variations. 

\begin{figure*}
    \includegraphics[width=6.8in]{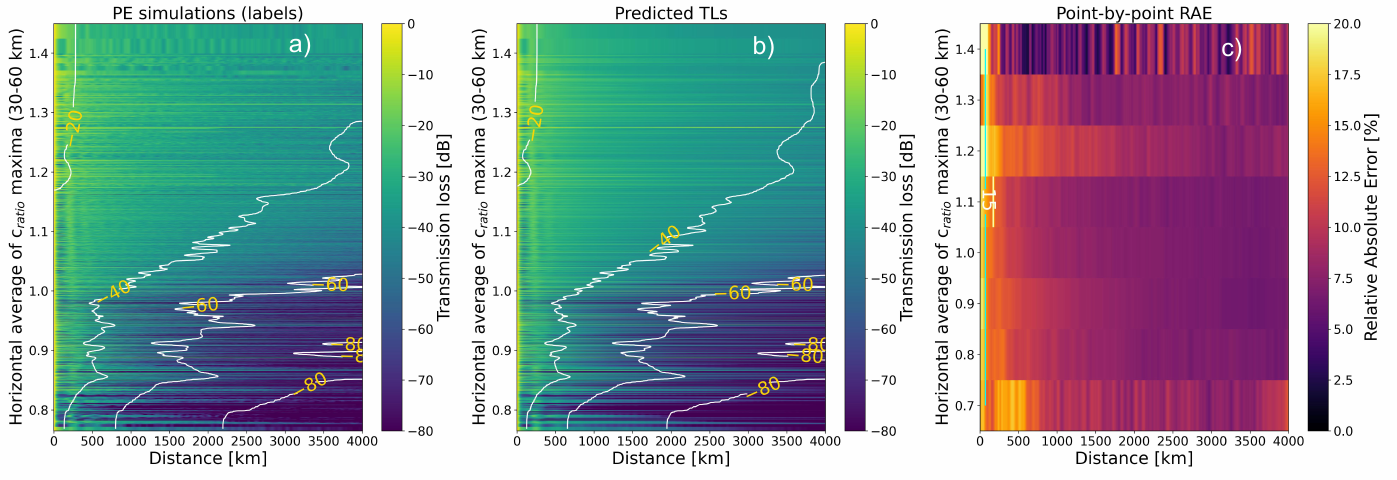}
    \caption{Comparison between $6,000$ PE simulations (panel a)) and ground-level TL predictions (panel b)) associated with a point-by-point error in percentage (panel c)).} 
    \label{fig:7} 
\end{figure*} 

\smallskip

We introduce two error metrics to further evaluate the performances: the Relative Absolute Error (MRAE) and the Root Mean Squared Error (RMSE) both averaged along the propagation path. The MRAE captures the difference in percentage between predictions $\hat{\ell}_f$ and PE simulations $\ell_f$: 
\begin{eqnarray}
\text{MRAE}=\frac{1}{D} \sum_{d=1}^{D}\left( \frac{ \rvert (\ell_{f})_{d} - (\hat{\ell_{f}})_{d} \rvert}{\rvert (\ell_{f})_{d} \rvert} \times 100 \right) \text{;}\quad D=800 \quad \text{points},
\label{eq:6}
\end{eqnarray}

where $d$ is the distance from the source. The MRAE is computed on decibel values coming from de-standardized TL. Such a metric penalizes errors close to the source equally \change[AJC]{as much at}{as at much} greater range. This prevents overweighting errors occurring at large distances, where the infrasounds are often very attenuated, so they do not allow signals to be detected above the background noise at IMS stations. The RMSE is computed to make comparisons with results published by \citeA{qb23}. The RMSE evaluates the difference between two de-standardized TLs directly in decibels.

\smallskip

\begin{figure}
    \includegraphics[width=3.8in]{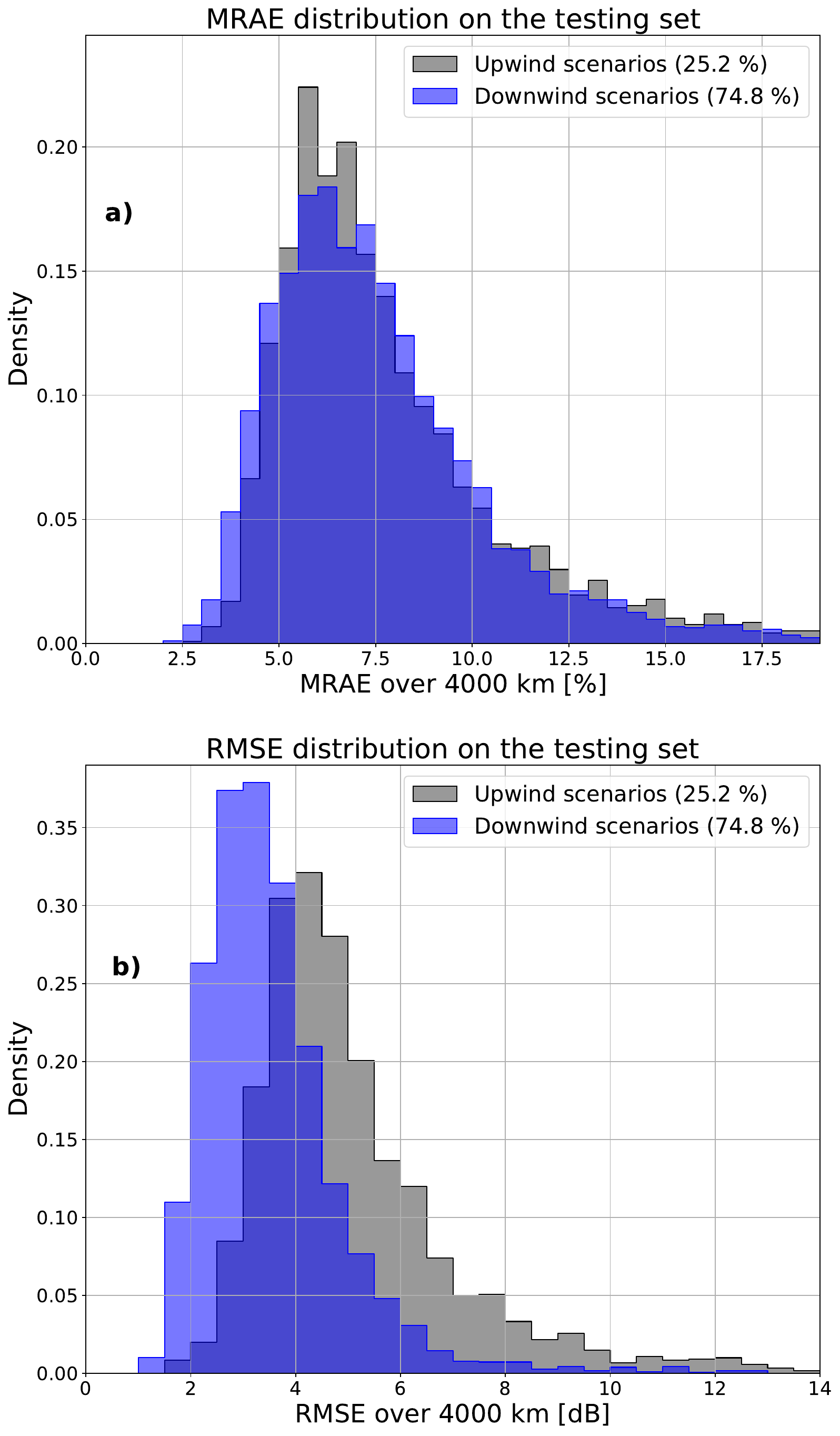}
    \caption{Distributions of the MRAE (panel a)) and of the RMSE (panel b)) along the $4,000$ km-long propagation paths for the $6,000$ samples of the testing dataset. A distinction is made between upwind and downwind scenarios.} 
    \label{fig:8} 
\end{figure}

For each sample of the testing dataset, the point-by-point relative absolute error (RAE) along the propagation path shows regions with errors reaching $15$\% in the first $250$ km (see Figure \ref{fig:7} panel c)). These higher error areas correspond to regions where most TL variation occurs within the first acoustic shadow zone and the first stratospheric return. The MRAE computed on the whole testing dataset follows a distribution peaking at an error of $6.2$\% (see Figure \ref{fig:8} panel a)). The median of the overall distribution is $7$\% with a $95$\%-percentile of $15,4$\%. 

The distribution of the RMSE has a mean of $4.3$ dB over the $4,000$ km-long propagation paths. A slight degradation of the performance is noted for upwind scenarios (see panel b) of Figure \ref{fig:8}). The predominant class of the error distribution is for downwind scenarios, with an RMSE between $2.5$ and $3.5$ dB of error, compared to $3.5$ to $4.5$ dB for the upwind scenarios. These results are consistent with \citeA{qb23}, which obtained an average RMSE of $5$ dB regardless of the initial wind conditions over $1,000$ km. This highlights the ability of our new model to maintain its performance against realistic atmospheric medium covering a long propagation range. 

\smallskip

As in \citeA{qb23}, our testing dataset analysis reveals a degradation of the performance with increasing frequency. Figure \ref{fig:9} shows the MRAE distributions obtained by $F_\theta(A_{z,d},f)$ for the five frequencies $f=0.1$, $0.2$, $0.4$, $0.8$ and $1.6$ Hz. The MRAE distributions restricted to the ground-level TL of higher frequencies ($f\ge0.8$ Hz) are wider than the one associated with the TL of lower frequencies (e.g., $14.4$\% of $95$\%-percentile at $0.1$ Hz and $20.3$\% at $1.6$ Hz). We explain this by the greater sensitivity of higher frequencies to small-scale variations, which are more challenging to predict with our current dataset size.

\begin{figure}
    \includegraphics[width=3.85in]{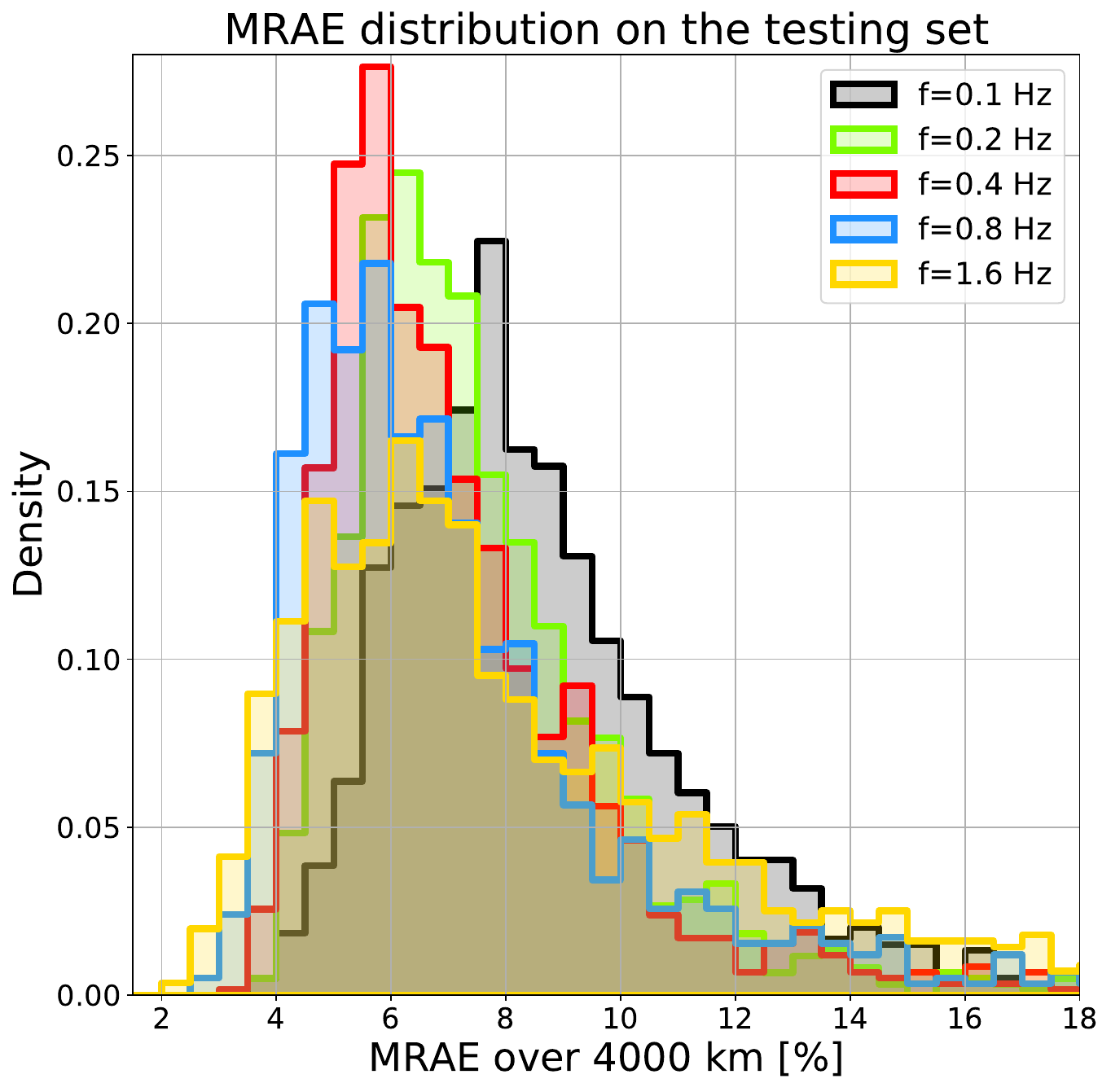}
    \caption{MRAE distributions along the $4,000$ km-long propagation paths for the five source frequencies. One can see the degradation of the performance with increasing frequency.} 
    \label{fig:9} 
\end{figure}

%%%%%%%%%%%%%%%%%%%%%%%%%%%%%%%%%%%%%%%%%%%%%%%%%%%%%%%%%%%%%%%%%%%%%%%%%%%%%%%%%%%%%%%%%%%%%%
\subsection{Performance assessment using the generalization dataset}
\label{results_gen}
To further evaluate the model performance, we construct a generalization dataset using $12$ sampling points previously left apart and keep all ten GW perturbation fields per atmospheric slice (see Section~\ref{training_validation_datasets} and blue lines in Figure \ref{fig:4}). The generalization samples correspond thus to never-before-encountered locations on the Earth sampled on the same date as the training/validation/testing data, with additional small-scale variation fields. As a result, the generalization dataset contains $K_{\text{gen}}=12 \times 8 \times 10 \times 2 \times 5=9.600$ samples. 

\smallskip

Analyzing  the variability of the atmospheric conditions present in the generalization dataset is important for a comprehensive performance evaluation of $F_\theta(A_{z,d},f)$. The comparison of the mean $c_\text{ratio}$ fields computed on the generalization and the training datasets reveals differences three times larger than the highest differences reached when making the same comparison on the testing and the training datasets, particularly in the thermosphere. We explain these observations by the use of a larger number of GW perturbation fields in the generalization dataset, whose effects are mainly visible in the upper layers of the atmosphere. See \ref{appendix_2} for more details on the distributions of minimal, average, and maximal values of $c_\text{ratio}$ in the troposphere, stratosphere, mesosphere, and thermosphere computed for the generalization and the training datasets. 

\smallskip

As the differences between the generalization and the training datasets is larger than between the testing and the training ones, we observe a slight degradation of the performance on the generalization data. We note a wider distribution of the MRAE over $4,000$ km and a larger median (of approximately $8.5$\% on the generalization samples instead of $7$\% on the testing ones). Figure \ref{fig:10} illustrates this result by comparing the MRAE distribution on the testing and the generalization datasets at $0.2$, $0.8$ and $1.6$ Hz. We can see a similar shift in the generalization distributions for all source frequencies. See \ref{appendix_3} for additional details. 

\begin{figure}
    \centering
    \includegraphics[width=6.8in]{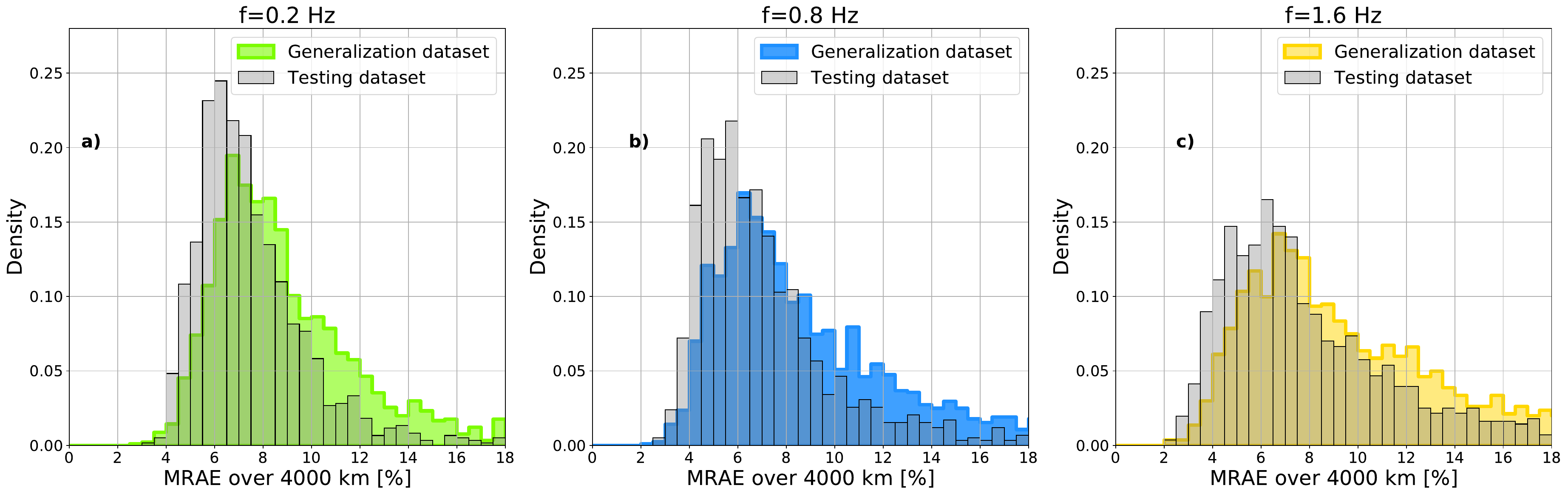}
    \caption{MRAE distributions at $0.2$, $0.8$ and $1.6$ Hz on the testing and the generalization datasets.} 
    \label{fig:10} 
\end{figure}

%%%%%%%%%%%%%%%%%%%%%%%%%%%%%%%%%%%%%%%%%%%%%%%%%%%%%%%%%%%%%%%%%%%%%%%%%%%%%%%%%%%%%%%%%%%%%%
%%%%%%%%%%%%%%%%%%%%%%%%%%%%%%%%%%%%%%%%%%%%%%%%%%%%%%%%%%%%%%%%%%%%%%%%%%%%%%%%%%%%%%%%%%%%%%
\section{Epistemic uncertainty and data sensitivity}
\label{section_5}
In the field of deep learning, quantifying the uncertainties is crucial for many applications, as in our case with the monitoring of explosive sources by the IMS infrasound network. In this section, we introduce two ways of quantifying the uncertainties associated with the predictions of the neural network, by distinguishing epistemic uncertainty from data uncertainty. 

\smallskip

The uncertainty related to the predictive process of a neural network is part of the "in-domain" category \cite{aa20}. It reflects the difficulty for a model to explain new samples from a data distribution assumed to be similar to the training data distribution due to a lack of intra-domain knowledge. The causes of such uncertainties may be directly related to the model architecture (epistemic) or to the data used to feed it. 

\smallskip

A way to quantify the uncertainties related to the model architecture is to use Bayesian methods. A Bayesian Neural Network $G_{\phi}(I)$ (\citeA{t89}; \citeA{wb91}) can predict a new sample $(I^*, O^*)$ that incorporates epistemic uncertainty by considering its parameters $\phi$ as random variables following a posterior distribution $p(\phi,(I,O))$ given observed samples $(I,O)$. This posterior distribution is based on prior belief on $\phi$ and on the use of Bayes’ theorem \cite{jg23}. The predictive posterior distribution for a new input $I^*$ is obtained by marginalizing over $\phi$, which implies integrating over all possible values of $\phi$ weighted by their posterior probabilities: 
\begin{eqnarray}
p(O^*|I^*,(I,O))=\int_{}^{}p(O^*|I^*,\phi) \times p(\phi,(I,O))d\phi
\label{eq:8}
\end{eqnarray}

However, calculating the posterior distribution $p(\phi,(I,O))$ and performing marginalization can be intractable for complex models such as neural networks. In this work, we use the Monte-Carlo Dropout method \cite{yg15} to approximate the predictive posterior distribution $p(O^*|I^*,(I,O))$, and indirectly the posterior distribution. $I$ corresponds to an input sample $(A_{z,d},f)$ and $O$ to a PE simulation $\ell_f$. We modify the architecture of $F_\theta(A_{z,d},f)$ so as to define its dropout layers as a set of random variables following a Bernoulli distribution. The units of these modified layers can be activated not only during training but also during a prediction. At the inference stage, the model is then no more deterministic but stochastic. The distribution associated with each new input sample $(A_{z,d},f)^*$ is obtained by averaging the predictions of the set formed by such Bayesian networks. This method allows us to provide a distribution of prediction for any new input while being inexpensive in terms of training time. 

\smallskip

The second source of uncertainty reflects the inherent noise of the data that feed the neural network. We use the Test-Time Augmentation method to quantify the sensitivity of $F_\theta(A_{z,d},f)$ to its input data. Several samples from each input $(A_{z,d},f)$ are generated by applying augmentation techniques and making predictions for each of them, which can then be averaged to obtain an output distribution. This method allows us to quantify the sensitivity of the model to the variations of its input data while being not resource-intensive because the architecture is not modified and does not require additional training samples. The augmentation applied on each input $(A_{z,d},f)$ implies superimposing ten realizations of the GW perturbation fields on the atmospheric slices $A_{z,d}$. Therefore, the simulated and predicted TLs are the result of the average of ten simulations/predictions originating from the same slice $A_{z,d}$. 

\smallskip

We estimate the epistemic and the data uncertainties on the generalization dataset. For each sample, the Monte-Carlo Dropout $F_\theta(A_{z,d},f)$ realizes $100$ predictions, whose means and standard deviations are calculated over $4,000$ km. The obtained epistemic uncertainty represented by the standard deviations is quite small, as shown in \ref{appendix_4}. The largest standard deviations (between $3$ dB and $5$ dB), are reached for upwind atmospheric scenarios with $f\ge0.8$ Hz. Above $0.8$ Hz, we note an increase in the model uncertainty in the first $250$ kilometers of propagation. At $1.6$ Hz, these areas spread over $500$ km for upwind cases with an average epistemic uncertainty of $4$ dB. All these results highlight the correlation between the worst predicted areas of the TL (see Section~\ref{results_test}) and the uncertainties related to the model architecture. Regarding data-related uncertainty, the comparison between the standard deviation obtained on the PE simulations and estimated by the model shows a low sensitivity of $F_\theta(A_{z,d},f)$ to its input data (see \ref{appendix_5}). The standard deviation associated with the predictions on the generalization dataset is on average $1.2$ dB, compared to $3.2$ dB for the corresponding PE simulations. This indicates that the uncertainty induced on the TL by different GW perturbation fields is not fully captured by $F_\theta(A_{z,d},f)$. The model acts as a low-pass filter, which is consistent with the observations obtained on the testing data (see Section~\ref{results_test}). 

\smallskip

Figure \ref{fig:11} details an example of ground-level TLs associated with a sample of the generalization dataset, all eight directions of propagation considered. The frequency is fixed at $0.2$ Hz. The predicted standard deviation (in orange) is estimated by considering both epistemic and data uncertainties, and is compared to the data uncertainty obtained on the PE simulations (in blue). One can see greater uncertainties for the directions of propagation associated with upwind atmospheric conditions (see panels c) and d)). Nevertheless, even in the directions where the total uncertainty is the largest, it does not recover all variations of the PE simulations. This is explained by the aforementioned disability of $F_\theta(A_{z,d},f)$ to capture all TL variations induced by the GW perturbation fields. Finally, panel i) presents the point-by-point relative absolute error averaged along the eight directions. It confirms the increase of error in the first $250$ km, reaching $15.6$\%. 

\begin{figure}
    \centering
    \includegraphics[width=5.85in]{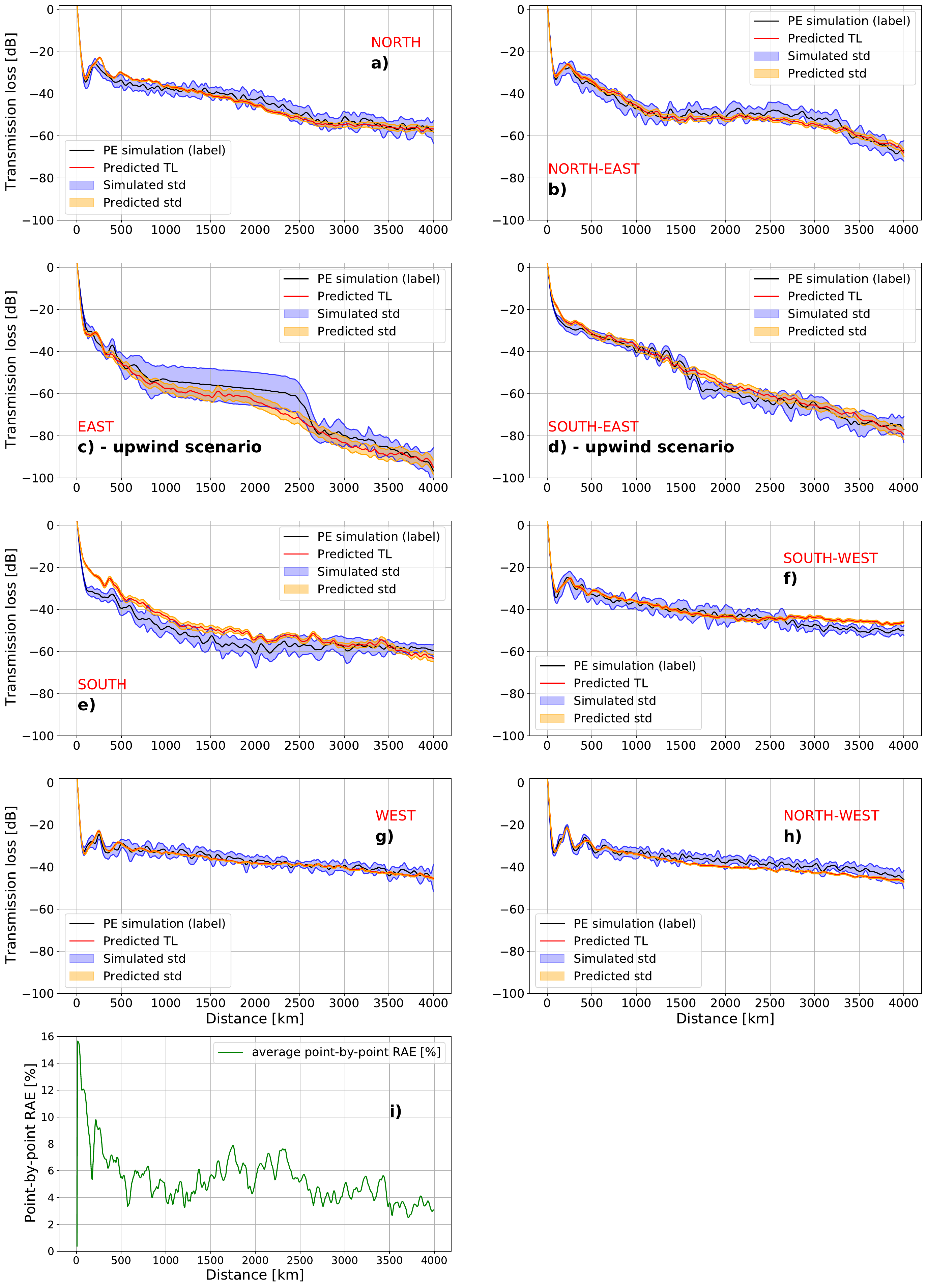}
    \caption{TLs associated with a sample of the generalization dataset at $f=0.2$ Hz, all directions of propagation considered. The predicted standard deviation (in orange) is computed by combining the epistemic and the data uncertainties. It is compared to the simulated data uncertainty obtained on the PEs (in blue). Panel i) represents the point-by-point RAE averaged along the eight directions of propagation.} 
    \label{fig:11} 
\end{figure}

%%%%%%%%%%%%%%%%%%%%%%%%%%%%%%%%%%%%%%%%%%%%%%%%%%%%%%%%%%%%%%%%%%%%%%%%%%%%%%%%%%%%%%%%%%%%%%
%%%%%%%%%%%%%%%%%%%%%%%%%%%%%%%%%%%%%%%%%%%%%%%%%%%%%%%%%%%%%%%%%%%%%%%%%%%%%%%%%%%%%%%%%%%%%%
\section{Discussion}
\label{section_6}

\subsection{General remarks}
The work presented so far is based on the assumption that sampling the Earth on a single date is sufficient to represent a large variability of the middle and upper atmosphere on a global scale. The date of January $15$ was chosen because it corresponds to a period of the year with extreme dynamical events, such as sudden stratospheric warmings \cite{mb21}. Moreover, in $2021$ that day, the winds reached zonal speeds of $-50$ $\text{m.s}^{-1}$ in the tropics and more than $150$ $\text{m.s}^{-1}$  at mid-latitudes in the northern hemisphere. Finally, the use of small-scale disturbance fields, such as the ones generated with the empirical GW model, increases the variability of the whole database, leading to a wide range of initial atmospheric conditions. 

\smallskip

\add[AJC]{In Section}~\ref{results_test}\add{, we introduced the spectral bias phenomenon as part of the explanation of the model}’\add{s "low pass filter" behavior. This bias could be mitigated either by pre-processing the input data in the frequency domain (e.g., standardizing the input source frequency, similarly to what was done with the input atmospheric slices), or by modifying the neural network architecture to better capture high-frequency features (e.g., by increasing the model}’\add{s depth and width, or by adding residual connections).}

\smallskip

Sections \ref{results_test} and \ref{results_gen} highlight the decrease of the performance of $F_\theta(A_{z,d},f)$ with increasing frequency. Two different databases were created to investigate this effect and try to limit the impact of the source frequency on the predictions. However, the obtained outcomes were not relevant. The neural network trained on the two new databases showed degraded overall performances compared to the approach presented so far, with similar degradation of performance with increasing frequency. See \ref{appendix_6} for additional details.

\smallskip

We perform various numbers of inferences to estimate the epistemic uncertainty using the Monte-Carlo Dropout method ($10$, $50$, $100$ as presented in Section~\ref{section_5}, and $150$). Each time, the average epistemic uncertainty remains below $2$ dB across all source frequencies. These uncertainties may seem low, but are quite expected since the Monte-Carlo Dropout method relies on the assumption that dropout layers introduce sufficient stochasticity during forward passes to effectively sample the model's parameters. However, this assumption may lead to a less accurate approximation of the Bayesian inference and resulting epistemic uncertainty. Such limitation can thus provide an additional explanation on the inability of $F_\theta(A_{z,d},f)$ to recover all variations of the PE simulations, as observed in Figure \ref{fig:11}.

%%%%%%%%%%%%%%%%%%%%%%%%%%%%%%%%%%%%%%%%%%%%%%%%%%%%%%%%%%%%%%%%%%%%%%%%%%%%%%%%%%%%%%%%%%%%%%
\subsection{Tonga case study: example of generalization in time, space and frequency}
\label{results_Tonga}
A direct application of our model is the near real-time prediction of TL maps around a source of interest. To emulate this use case, we choose as source the Tonga volcano, which erupted on January 15, $2022$. This event is a reference case in the infrasound community and has been widely documented (\citeA{rm22}; \citeA{jv22}; \citeA{ap22}). The eruption generated waves that were detected by all the operational IMS infrasound stations. 

\smallskip

We collect $360$ atmospheric slices around the Tonga volcano to form a new generalization dataset; the \textit{Tonga-set}. Like the previous generalization dataset presented in Section~\ref{results_gen}}, these samples correspond to never-before-encountered locations on the Earth. In addition, the date of sampling is different. The mean $c_\text{ratio}$ fields computed on the training dataset and the \textit{Tonga-set} reveal regions with a difference up to $0.1$ of $c_\text{ratio}$ above the mesopause (see Figure \ref{fig:12}). This is six times larger than the highest difference reached when making the same comparison between the training and the previous generalization datasets. The distributions of minimal, average, and maximal values of $c_\text{ratio}$ in the troposphere, stratosphere, mesosphere, and thermosphere give more details on these differences (see \ref{appendix_7}). As an illustration, the training dataset contains scenarios with a minimal $c_\text{ratio}$ between $0.65$ and $0.73$ in the troposphere and between $0.4$ and $0.6$ in the stratosphere, which are situations totally absent in the \textit{Tonga-set}. On the contrary, the proportion of scenarios with a minimal $c_\text{ratio}$ between $0.7$ and $0.8$ in the stratosphere or with a mean $c_\text{ratio}$ between $1.0$ and $1.15$ in the thermosphere is over-represented in the \textit{Tonga-set}. Similarly, the \textit{Tonga-set} over-represents samples with a $c_\text{ratio}$ maximum of $1.6$ above $90$ km altitude. Such differences impact the generalization capabilities of the neural network, which depends on the similarities between the training and the evaluation datasets. 

\smallskip

\begin{figure}
    \includegraphics[width=3.5in]{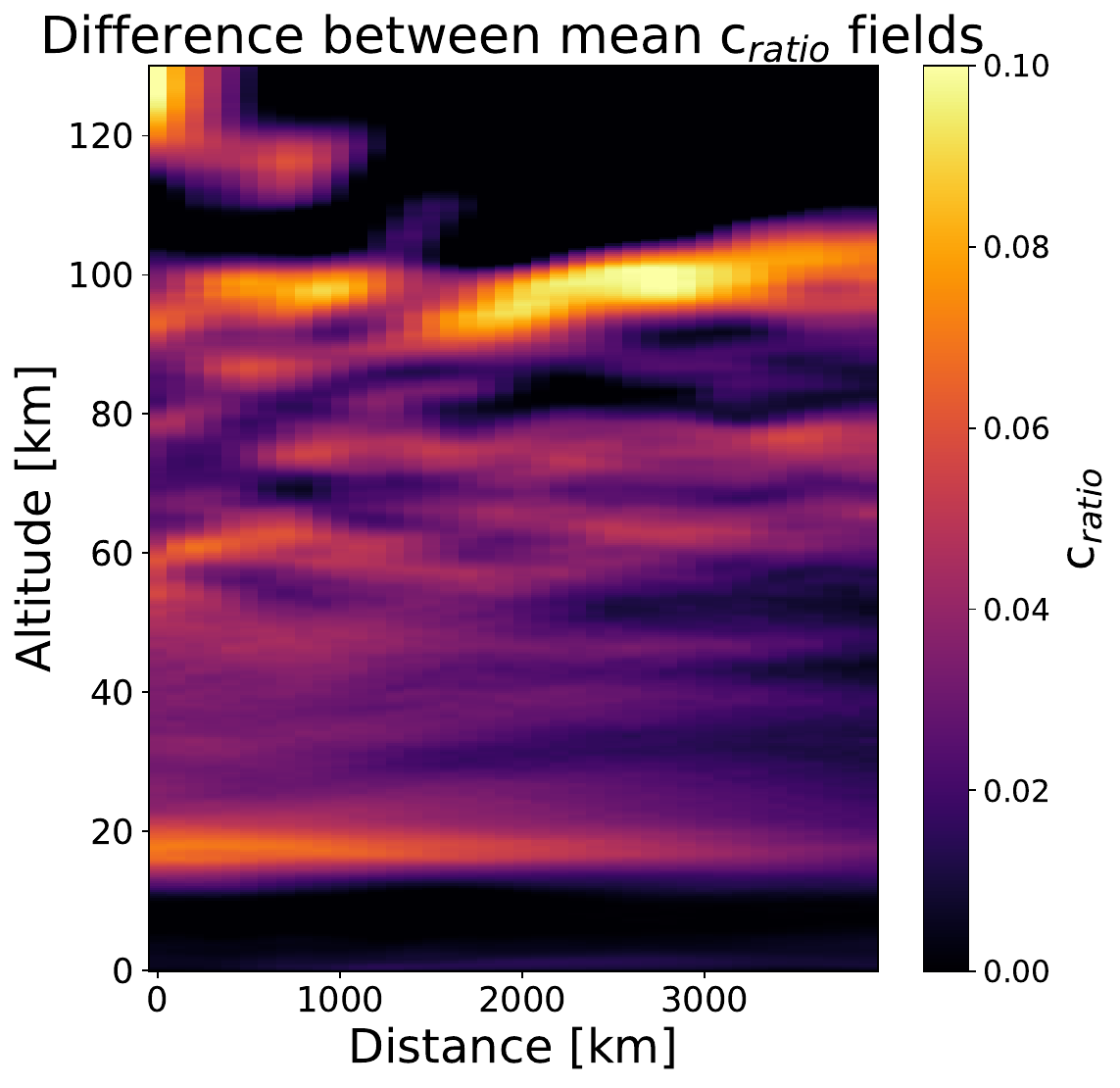}
    \caption{Differences between the mean $c_\text{ratio}$ fields computed on all the samples of the training dataset and the \textit{Tonga-set}.}
    \label{fig:12} 
\end{figure}  

At $01:00:00$ UTC on January $15$, $2022$, atmospheric models predict regions with both downwind and upwind propagation conditions. Stratospheric guiding is visible towards the west of the Tonga volcano (200$^\circ$--315$^\circ$ azimuth) with mean $c_\text{ratio}$ reaching $1.2$ at $50$ km altitude. On the contrary, for eastward propagation, the mean $c_\text{ratio}$ reaches $0.6$ in the stratosphere. At lower altitudes, we observe the existence of a weak tropospheric guide at the top of the boundary layer in all directions. In addition, some guides exist just above the mesopause, mainly toward $135^\circ$ azimuth. 

\smallskip

\begin{figure}
    \includegraphics[width=3.2in]{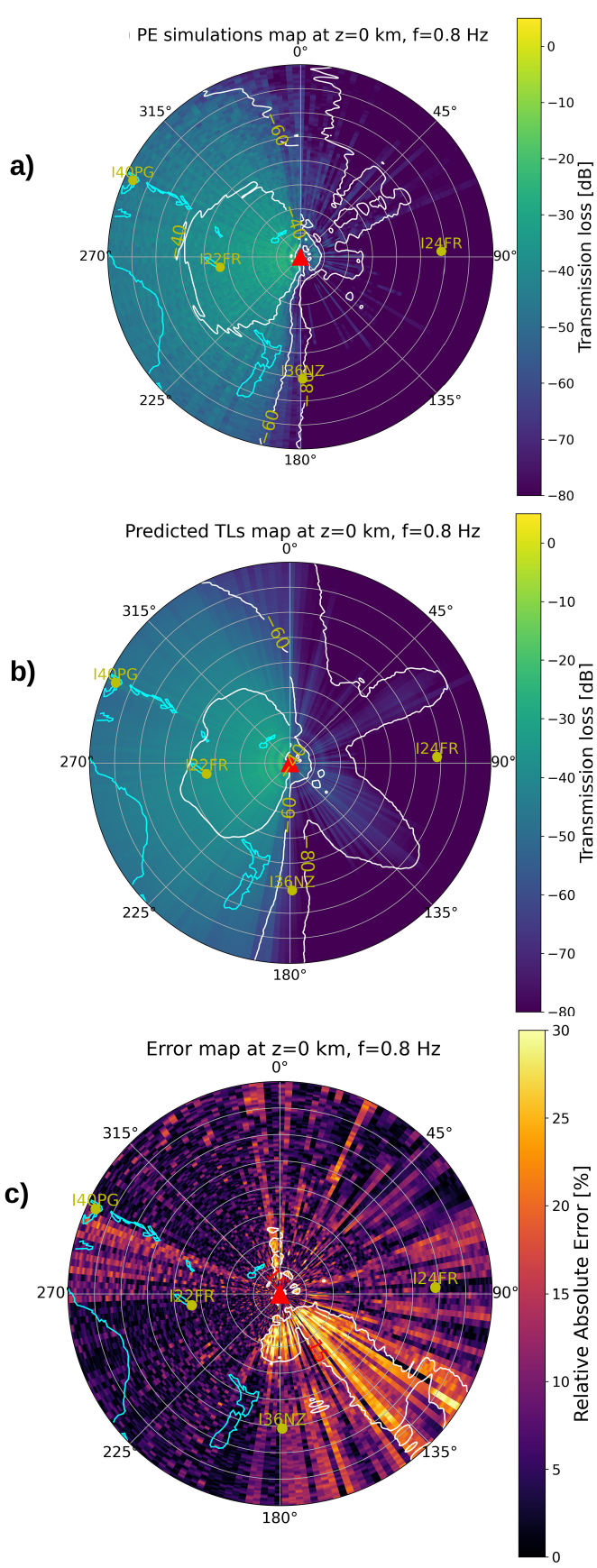}
    \caption{Simulated and predicted TL maps around the Tonga volcano (panels a) and b)), and point-by-point RAE in percentage (panel c)) at $f=0.8$ Hz. \add[AJC]{The four IMS stations located within a $4,000$ km radius of the Tonga volcano are marked with red dots on the maps (IS22, IS24, IS36 and IS40).}} 
    \label{fig:13} 
\end{figure}

At the inference stage and with a batch of $360$ samples, $F_\theta(A_{z,d},f)$ predicts the $4,000$ km-long ground-level TLs all around the volcano in less than $0.3$ seconds, regardless the frequency (Dell Inc. Intel(R) Core(TM) i$9$-$13900$ $48$ CPUs $77.8$ GB RAM on RedHat $9.5$). The error between the PE simulations and the predicted TLs is quantified using the point-by-point RAE. Despite the differences observed between the training dataset and the \textit{Tonga-set}, the median MRAE computed on this last set is $7.9$\% only. This illustrates the ability of the surrogate model to accurately predict TL maps for an event absent of its training and validation datasets. Figure \ref{fig:13} shows simulated and predicted TL maps (panels a) and b)) as well as the point-by-point RAE (panel c)) at $f=0.8$ Hz. The TL reaches $-125.3$ dB eastward, where infrasound waves are not refracted to the ground in the stratosphere but propagate into the thermosphere. This direction is associated with a global increase of the errors, particularly towards $135^\circ$ azimuth where a waveguide exists just above the mesopause. These areas with higher errors might be due to the aforementioned differences observed above the mesopause when comparing the training dataset and the \textit{Tonga-set}. 

\ref{appendix_8} shows the distribution of the MRAE as a function of frequency and mean $c_\text{ratio}$ in the troposphere, stratosphere, mesosphere, and thermosphere. This confirms poorer performances in upwind cases and with increasing frequency. At $1.6$ Hz, the $95$\%-percentile of the MRAE raises up to $26.4$\% which is by far the highest error we observe. Such sensitivity to downwind/upwind conditions is greater than what was observed previously for the testing dataset (see Section~\ref{results_test}). We explain this by the lack of representation of scenarios with a minimal stratospheric $c_\text{ratio}$ between $0.7$ and $0.8$ in the training dataset compared to the \textit{Tonga-set}, preventing the surrogate model from correctly learning how to predict the TL in that situation. In the absence of stratospheric guides (upwind), infrasound waves reach the layers above the stratosphere. This corresponds again to a situation where the training dataset and the \textit{Tonga-set} differ significantly, particularly above the mesopause. 

\smallskip

\begin{figure}
    \includegraphics[width=3.25in]{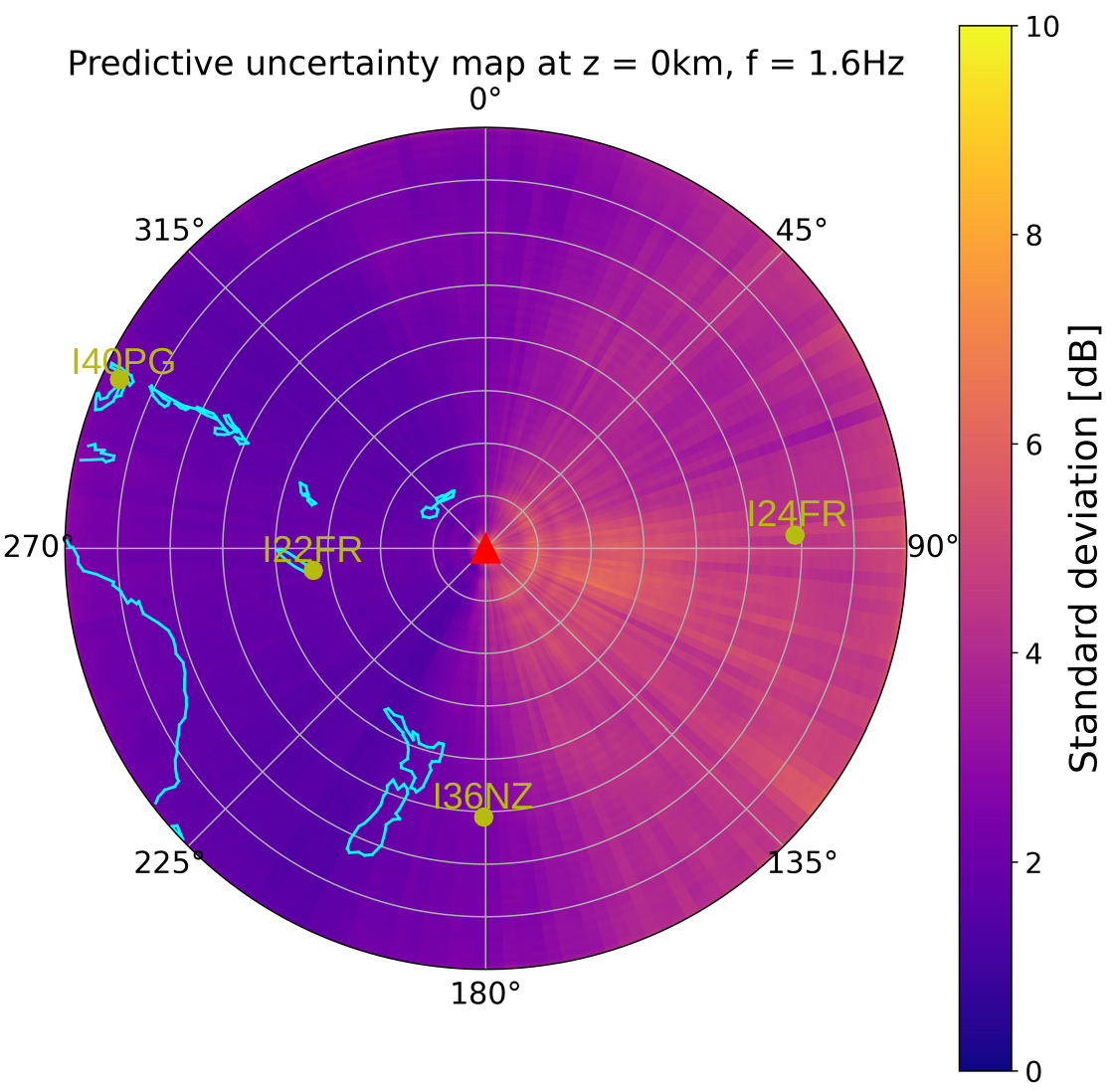}
    \caption{Epistemic and data uncertainty map (standard deviation in dB) of the predictions realized at $1.6$ Hz.} 
    \label{fig:14} 
\end{figure}

For the $360$ predictions at each frequency, the epistemic and the data uncertainties are quantified using a combination of the Monte-Carlo Dropout method with the Test-Time Augmentation technique. The total uncertainty of a given prediction corresponds to the standard deviation obtained from an initial atmospheric slice disturbed by ten GW perturbation fields and whose associated TLs are simulated/predicted ten times by the set of Monte-Carlo-Dropout-$F_\theta(A_{z,d},f)$s. The total uncertainty increases with increasing frequency and for upwind conditions. Figure \ref{fig:14} shows the total uncertainty map obtained by $F_\theta(A_{z,d},f)$ at $f=1.6$ Hz. The standard deviation reaches more than $10$ dB eastward, where the $95$\%-percentile of the MRAE raises up to $26.4$\%. This highlights again that high uncertainties are estimated in regions of large errors. 

\smallskip

To further deepen the generalization capabilities of $F_\theta(A_{z,d},f)$, we therefore evaluate it on the \textit{Tonga-set} at never-before-encountered source frequencies. We use the same set of $360$ atmospheric slices $A_{z,d}$ as input, but at the following frequencies $f$: $0.3$, $0.6$, $1.0$, $1.2$ and $1.4$ Hz. As previously, \ref{appendix_9} displays the distributions of the MRAE according to these frequencies and to the mean $c_\text{ratio}$ in the troposphere, stratosphere, mesosphere, and thermosphere. The strong correlation between the error distributions obtained on the set of new frequencies compared with the original distributions is highlighted. For each distribution, the errors remain in the same order of magnitude. Higher errors (between $20$\% and $30$\%) spread on the same range of mean $c_\text{ratio}$ values (mainly for upwind cases). However, the increase of error with increasing frequency is now visible as early as $f$ becomes greater or equal to $0.6$ Hz. 

\smallskip

The \textit{Tonga-set} study thus shows the overall ability of $F_\theta(A_{z,d},f)$ to generalize to new atmospheric model regimes, new dates, and new source frequencies.

%%%%%%%%%%%%%%%%%%%%%%%%%%%%%%%%%%%%%%%%%%%%%%%%%%%%%%%%%%%%%%%%%%%%%%%%%%%%%%%%%%%%%%%%%%%%%%
%%%%%%%%%%%%%%%%%%%%%%%%%%%%%%%%%%%%%%%%%%%%%%%%%%%%%%%%%%%%%%%%%%%%%%%%%%%%%%%%%%%%%%%%%%%%%%
\section{Conclusion}
\label{section_7}
In this study, we have shown the ability of a deep learning algorithm to estimate accurately and almost instantaneously infrasound ground-level TLs over a distance of $4,000$ km; at $f \in [0.1, 0.2, 0.4, 0.8, 1.6]$ Hz and for various initial atmospheric conditions $A_{z,d}$. These conditions are modeled using horizontal wind speed and temperature values extracted from the WACCM operational product, to which range-dependent fine-scale GW perturbation fields have been superimposed. 

\smallskip

The neural network is a Convolutional Recurrent Neural Network inspired by the CNN developed in \citeA{qb23}. Our addition of recurrent layers after the convolution layers allows the model to extract both spatially local and range-dependent features embedded in realistic atmospheric models. The neural network was trained and validated on $42,000$ samples $(A_{z,d},f)$ to minimize the RMSE between predicted and simulated ground-level TL. Once done, the model estimates a TL in approximately $0.045$ seconds, regardless the frequency. This constitutes a major improvement compared to other numerical propagation modeling tools\add[AJC]{ such as parabolic equations solvers, with a computation time saving between three and four orders of magnitude}. 

\smallskip

The performance of the model was evaluated on $6,000$ testing samples. A median MRAE of $7$\% and a mean RMSE of $4.3$ dB were obtained, all source frequencies considered. The robustness of the model to the atmospheric conditions was highlighted, as well as its "low-pass filter" property. An overall decrease in performance with increasing frequency has also been noticed. All these results are consistent with the ones obtained by \citeA{qb23}. However, as in \citeA{qb23}, we cannot assert that our model respects the inherent range-dependence of the atmospheric models. Indeed, despite the use of recurrent layers, the filters in the convolutional layer of $F_\theta(A_{z,d},f)$ use local information before and after each given distance to encode spatial features.

\smallskip

$F_\theta(A_{z,d},f)$ was then evaluated on atmospheric conditions that differ more from the training one. Generalization data correspond to atmospheric slices built from previously randomly drawn points on the Earth, all eight propagation directions considered. Twice as many GW perturbation fields were applied to these slices to further increase their variability compared to the training data. As expected, the model reached poorer performances on the generalization samples than on the testing ones (median MRAE $8.5$\%). In order to improve the generalization capabilities, one could further increase the dataset size and its variability. This could be done by sampling the Earth's atmosphere at different periods of the year and at different years.

\smallskip

The expected TL was simulated using the PE solver \textit{ePape}. The Sutherland-Bass coefficients used in \textit{ePape} allow the inclusion of the absorption of acoustic energy by the atmosphere. However, these coefficients are unable to account for second-order effects due to changes in viscosity and specific heat ratio with deviations in the atmospheric composition above $90$ km altitude \cite{ls04}. Moreover, it is important to note that the PE method is a linear method that cannot model non-linear propagation effects which become more prominent at higher altitudes (above the mesopause). To limit computational cost, the propagation problem was simplified in Cartesian coordinates, assuming flat terrain with infinite ground impedance at sea level, and by pre-selecting only five frequencies. Given these limitations which may introduce bias in the simulations, numerical explorations with fully resolved time- and range-dependent wave propagation techniques accounting for non-linear propagation effects would provide more realistic simulation results.

\smallskip

The generalization data was used to estimate epistemic and data-related uncertainties. This revealed an increase in the uncertainties with increasing frequency, particularly for upwind atmospheric scenarios. A similar result was obtained when studying the input data sensitivity. The combination of the two uncertainties partly recovers the whole variability of the simulated TL considering multiple GW perturbation fields. There are also other techniques that could be explored to quantify the uncertainties in more detail, such as deep ensemble approaches \cite{bl17}. 

\smallskip

The generalization capabilities of $F_\theta(A_{z,d},f)$ were further demonstrated with an estimate of near real-time predictions of TL maps for infrasound generated by the eruption of the Tonga volcano. The slices extracted around the volcano represented never-before-encountered atmospheric conditions in terms of location and date. The median MRAE value was $7.9$\%, all source frequencies considered. We highlighted the consistent increase of the estimated uncertainties in these regions. Finally, we demonstrated the ability of the model to predict the TL maps in all directions at five new source frequencies. The obtained error distributions were slightly larger than those of the original frequencies. This study represents a first step towards the near real-time assessment of the global detection capabilities of the IMS infrasound network. \change[AJC]{This could be obtained using, e.g., the Bayesian approach developed by Blom et al. (2023) or by applying the method of Le Pichon et al. (2012) while estimating more accurately the TL.}{This could be obtained by using the proposed architecture to refine statistical models for transmission loss predictions as implemented in the Bayesian approach developed by} \citeA{pb23} \add{or by estimating more accurately TLs from a global source grid to all receivers} \cite{jv12}. \add{For a given source yields, the machine learning-based TLs could be used to predict the expected amplitude at a specific frequency for a set of IMS infrasound stations around the Tonga volcano (e.g., stations IS22, IS24, IS36, and IS40, as shown in Figures}~\ref{fig:13} and \ref{fig:14}\add{). By computing the signal-to-noise ratio based on the expected noise levels at these stations and defining a threshold, it would then be possible to determine which stations are capable of detecting the signals.}

\smallskip

Suggested future works include the specialization of the neural network on a regional scale using fine-tuning. An expected outcome would be an increase in performance when predicting TL on global and regional reference events (e.g., Tonga volcano eruption, Lebanese explosion in Beirut -- \citeA{cp21}, Finnish explosions Hukkakero -- \citeA{ev23}, Negev desert controlled explosion -- \citeA{dv13}). \add[AJC]{Increasing the amount of input data could further enhance the neural network performance.} \ref{appendix_10} \add{shows promising preliminary results, obtained by doubling the size of the training database by sampling the Earth’s atmosphere both on January $15$, $2021$, and August $15$, $2021$.} One could also compute the adjoint of the network in the context of atmospheric data assimilation to improve numerical weather prediction \cite{pl24}. This application takes direct advantage of the nature of the neural networks, since numerical propagation methods such as PE solvers do not allow the determination of all the partial derivatives needed to calculate the adjoint without being computationally expensive. Finally, a topic of interest for future elaboration is the modeling of two-dimensional outputs, allowing for predicting the TL also for receivers at altitude. The predicted two-dimensional TL could be exploited for further atmospheric investigations using stratospheric balloon observations of large explosive sources (\citeA{ap22}; \citeA{sa23}; \citeA{db21}; \citeA{es23}).

%%%%%%%%%%%%%%%%%%%%%%%%%%%%%%%%%%%%%%%%%%%%%%%
%% Optional Appendices go here
%%%%%%%%%%%%%%%%%%%%%%%%%%%%%%%%%%%%%%%%%%%%%%%
\clearpage
\appendix
\section{Construction of two-dimensional range-dependent GW perturbation fields}
\label{appendix_1}
The vertical GW spectrum as a function of the vertical wavenumber $m$ essentially consists in two regimes. The first part, $m < m^*$, is the non-saturated part driven by the GW source, with an \remove[AJC]{exponential} increase until the maximum of spectral energy at $m=m^*$\add[AJC]{, where $m^*$ is called the dominant wavenumber}. At larger wavenumbers, $m > m^*$, this is the saturated part of the spectrum \cite{sm87} with a characteristic $-3$ slope in logarithmic scale (see Figure 1 of \citeA{cg93}). Typical values for $m^*$ are a few to about $10$ kilometers as documented in \citeA{cg93}, and in the literature more generally (\citeA{al95}; \citeA{ho94}; \citeA{ch18}).

\smallskip

The method used \change[AJC]{in the current contribution}{throughout this study} consists in using the analytical expression of the vertical GW spectrum presented in \citeA{cg93} \remove[AJC]{(their equation $7$)} with the vertical profile of $m^*$  \remove[AJC]{(their equation $43$)} to build vertical profiles of GW perturbations for a given set of $10$-km thick overlapping atmospheric layers ($50\%$ overlap). \add[AJC]{The analytical expression of the vertical GW spectrum is given by} \citeA{cg93} (their equation $7$):
\begin{equation}
F_{u}(m) = \left\{
    \begin{array}{ll}
        2\pi\frac{\alpha N^{2}}{m_{*}^{3}}\left( \frac{m}{m_{*}} \right)^{S} & m \le m_{*} \\
        2\pi\frac{\alpha N^{2}}{m^{3}} & m_{*} \le m \le m_{b} \\
        2\pi\frac{\alpha N^{2}}{m_{b}^{3}}\left( \frac{m_{b}}{m} \right)^{\frac{5}{3}} & m_{b} \le m;
    \end{array}
\right.
\end{equation}

\add[AJC]{where the non-saturated regime ($m \le m_{*}$), the saturated regime ($m_{*} \le m \le m_{b}$) and the turbulence regime ($m_{b} \le m$) are represented. The term $m_{b}$ is defined as the buoyancy wavenumber identifying the transition between saturated GW and turbulence, and $N$ as the buoyancy frequency reflecting the vertical stability of the atmosphere. The term $\alpha$ is a constant below $1$ accounting for superposition effects when multiple waves interact, generating instabilities which lower the threshold of wave saturation \cite{cg93}. 

Regarding the vertical profile of $m^*$, it is given by} \citeA{cg93} (their equation 43):
\begin{equation}
    m_{*} \sim \text{exp}\left[-\frac{z}{(q+s)H)} \right];
\end{equation}

\add[AJC]{where $z$ is the altitude, $H$ the atmospheric scale height, and $s$ and $-q$ the spectral indices of the $m$-spectrum in the source regime and the saturation regime, which are key parameters that influence both the shape and the magnitude of the horizontal wavenumber spectrum.

For each of the vertical GW spectra that are modeled for a set of altitudes z, the inverse Fourier transform is derived given a randomly chosen phase. This provides vertical perturbation profiles, which respectively characterize each of the atmospheric layers centered on each $z$. Then, all these vertical profiles are weighted and linearly combined to produce a full perturbation profile for the whole atmosphere, retaining perturbation amplitudes characteristic of each altitude.} The Gaussian weights used in the combination are centered at mid-layers and normalized to ensure the total weight addressed to a given atmospheric layer is $100\%$, thus avoiding an excess of redistributed energy in any of the layers. This method is applied for a given number of randomly chosen phases, producing an ensemble of several realizations of GW perturbation profiles \add[AJC]{as shown in Figure} \ref{fig:2} \add{panel a).}

\smallskip

In the last stage, randomly chosen vertical GW profiles \add[AJC]{(among those derived above)} are horizontally and linearly combined using a horizontal correlation length documented in \citeA{cg93} (their section $4$ and Table $4$). The final product is a two-dimensional range-dependent GW perturbation field as shown in Figure \ref{fig:2} panel b). Such fields are superimposed to the $c_\text{ratio}$ slices \add[AJC]{of our database.}

%%%%%%%%%%%%%%%%%%%%%%%%%%%%%%%%%%%%%%%%%%%%%%%
\clearpage
\section{Details on the comparisons between the training and the generalization datasets}
\label{appendix_2}
\begin{figure}
    \centering
    \includegraphics[width=6.5in]{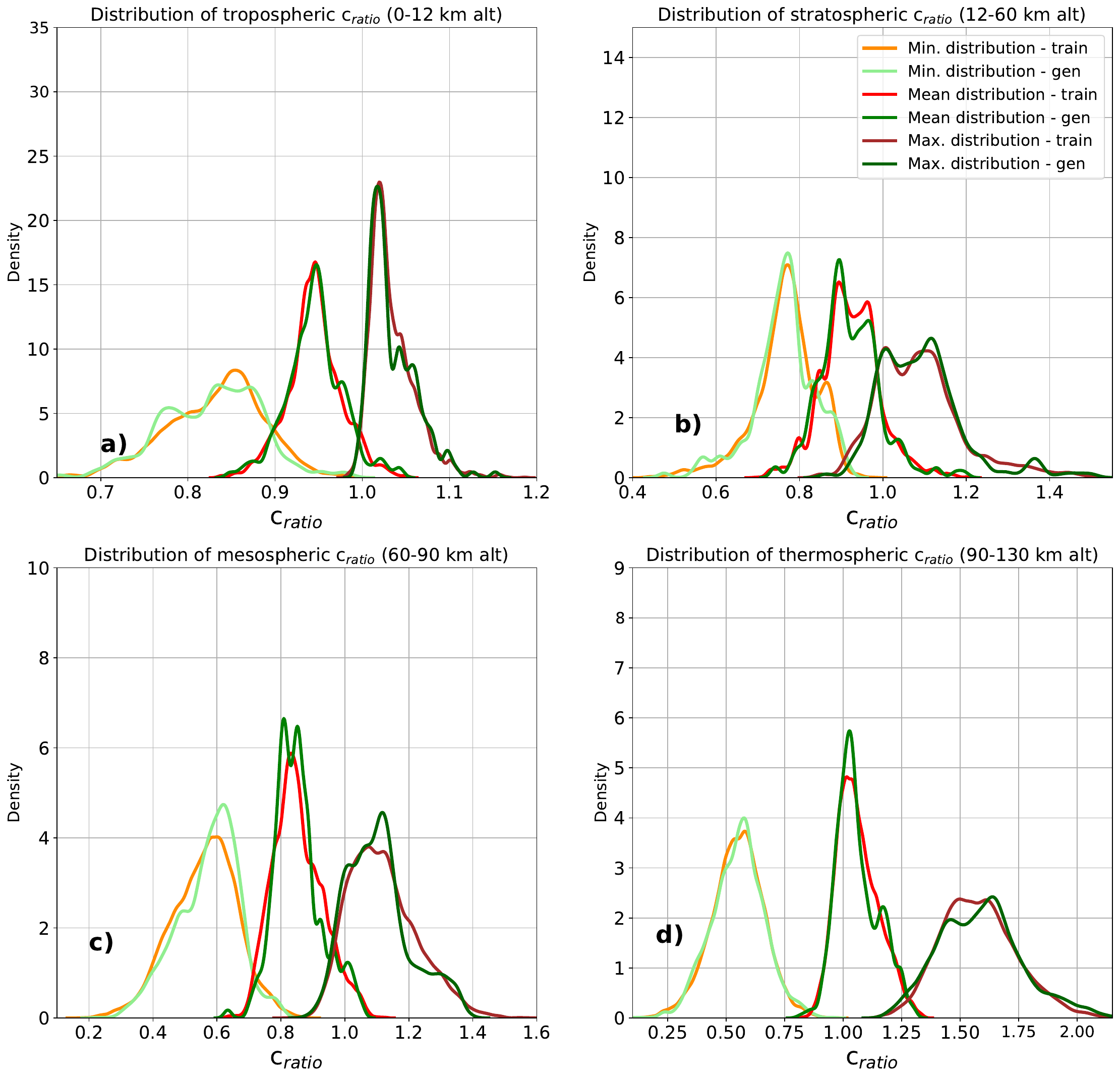}
    \caption{Distributions of minimal, average, and maximal values of $c_\text{ratio}$ in the troposphere, stratosphere, mesosphere, and thermosphere computed on the training and the generalization datasets.} 
    \label{fig:15} 
\end{figure}

This appendix aims at detailing the comparison realized between the training and the generalization datasets presented in Section~\ref{results_gen}. Each panel of Figure \ref{fig:15} focuses on a specific atmospheric layer. The altitudes of each layer are defined as presented in Section~\ref{section_2}, with the troposphere being defined between $0$ and $12$ km, the stratosphere between $12$ and $60$ km, the mesosphere between $60$ and $90$ km, and the thermosphere between $90$ and $130$ km altitude. For each layer, the minimal, average, and maximal $c_\text{ratio}$ distributions are plotted for both the training and the generalization datasets. As an example, the minimal distribution is obtained by distributing all the minimal values of $c_\text{ratio}$ among twenty bins of $c_\text{ratio}$ ranging from $0.1$ to $2.0$ with a step of $0.1$. As the training and the generalization datasets contain different numbers of cases, all results are plotted as a density. 

\smallskip

For each layer, we observe small departures of the distributions of the generalization atmospheric slices $A_{z,d}$ compared to the ones obtained on the training dataset. These departures can take the form of some situations slightly over-represented in the generalization dataset, like in the mesosphere  with a minimal $c_\text{ratio}$ of $0.6$ or a maximal $c_\text{ratio}$ of $1.15$. We explain this by the differences induced in the way we extract the generalization slices compared to the training ones. Generalization samples correspond to $12$ sampling points randomly set apart from the $162$ initial ones, associated with ten GW perturbation fields (see blue lines in Figure \ref{fig:4}). Generalization data correspond thus to never-before-encountered locations on the Earth sampled on the same date as the training ones, with additional small-scale variations. However, it is important to notice that the variability induced by the GW perturbation fields remains minor and is mainly visible above the stratosphere (see Figure \ref{fig:2} panel b)). The minimal and maximal zonal wind speed disturbances induced by these fields are on average $-25/25$ $\text{m.s}^{-1}$.

%%%%%%%%%%%%%%%%%%%%%%%%%%%%%%%%%%%%%%%%%%%%%%%
\clearpage
\section{Comparison of the MRAE distributions obtained on the testing and the generalization datasets}
\label{appendix_3}
\begin{figure}
    \centering
    \includegraphics[width=5.85in]{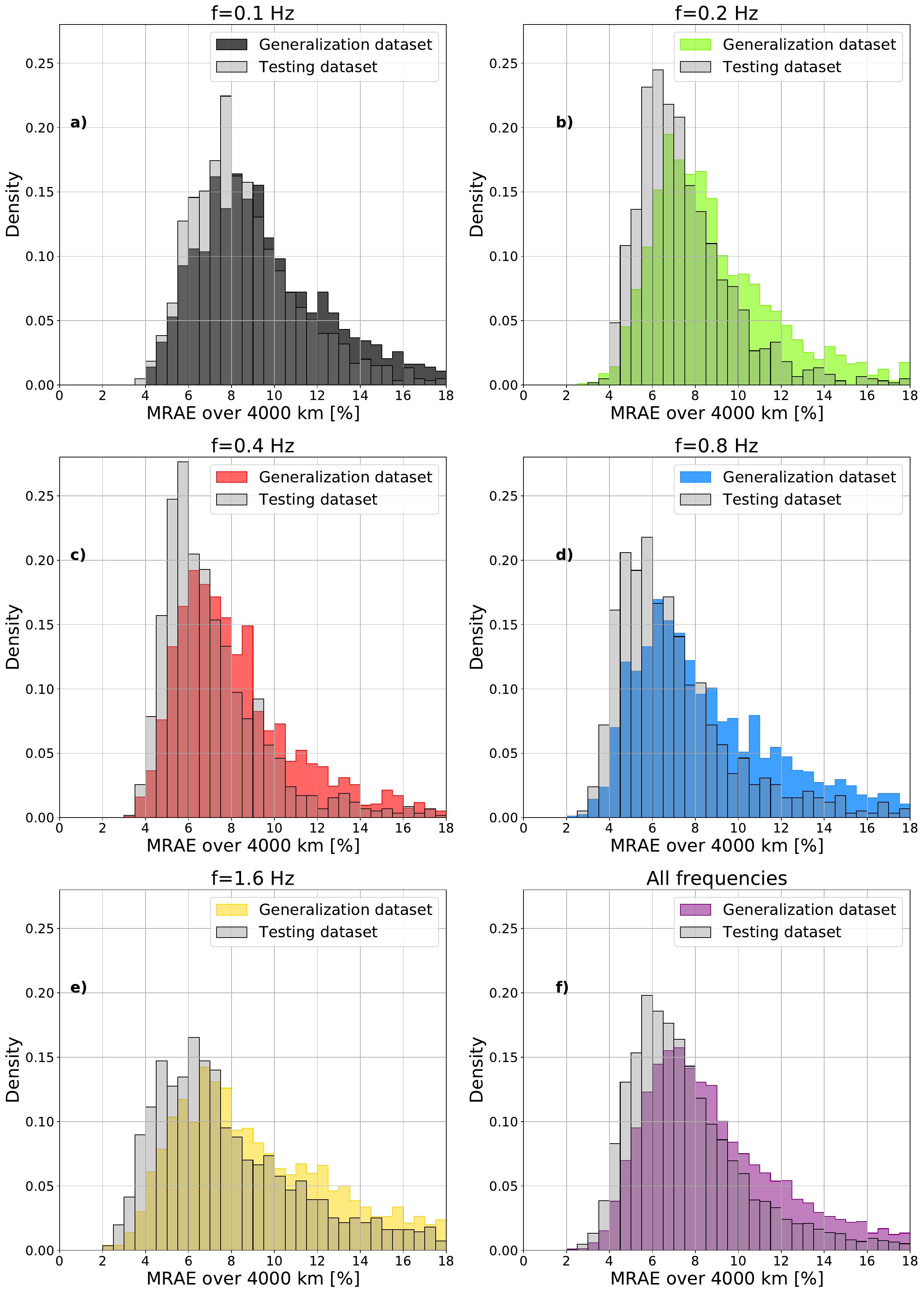}
    \caption{MRAE distributions at $0.1$, $0.2$, $0.4$, $0.8$ and $1.6$ Hz, as well as at all considered source frequencies, computed on the testing and the generalization datasets.} 
    \label{fig:16} 
\end{figure}

In Section~\ref{results_test}, we discussed the differences between the testing and the generalization performances. The current Appendix provides a more detailed view on that topic, by showing the MRAE distributions obtained on these datasets according to the five source frequencies as well as the overall distribution (see panel d) of Figure \ref{fig:16}). As observed on the testing dataset, the generalization performances decrease with increasing frequency. In addition to that, we observe for each frequency $f$ a slight shift of the generalization distributions towards larger errors, even if this shift is only $1$\% approximately regardless the frequency. This demonstrates the robustness of the model's generalization capabilities to the increase in frequency. 

%%%%%%%%%%%%%%%%%%%%%%%%%%%%%%%%%%%%%%%%%%%%%%%
\clearpage
\section{Epistemic uncertainty on the generalization dataset}
\label{appendix_4}
\begin{figure}
    \centering
    \includegraphics[width=6.4in]{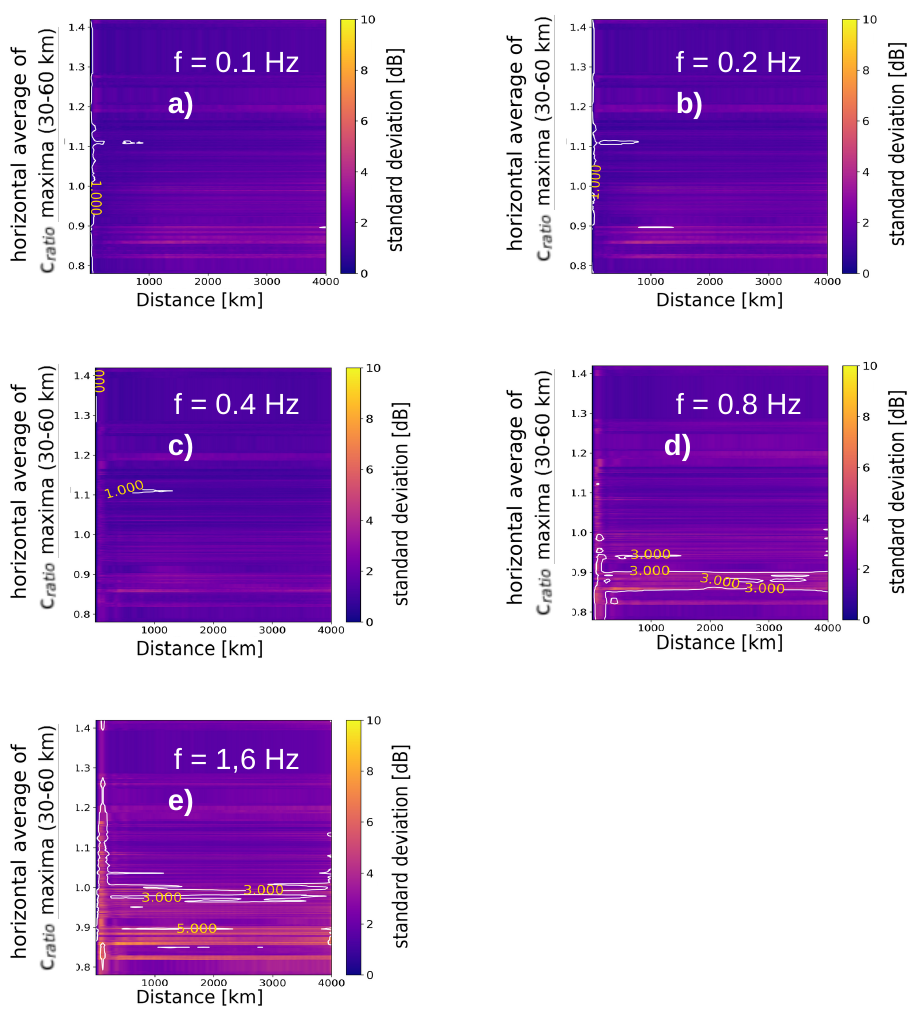}
    \caption{Uncertainty related to the surrogate model architecture, quantified on the generalization dataset for the five source frequencies using the Monte-Carlo Dropout method.} 
    \label{fig:17} 
\end{figure}

This appendix details the epistemic uncertainty obtained on the generalization dataset, for the five source frequencies. We use a Bayesian framework to quantify it, namely the Monte-Carlo Dropout method. As mentioned in Section~\ref{section_5}, the standard deviation associated with all the predictions of the generalization dataset is relatively small, with on average $1.98$ dB for all source frequencies. Figure \ref{fig:17} highlights the increase of epistemic uncertainty with increasing frequency; with on average $1.6$ dB for $f\le0.4$ Hz up to $2.18$ dB at $0.8$ Hz and $2.8$ dB at $1.6$ Hz. Finally, we observe an increase of epistemic uncertainty in the first $250$ km from the source, particularly at higher frequencies. All of this demonstrates the correlation existing between the higher uncertainty estimates and the areas with higher predictive errors (see Section~\ref{results_test}). 

%%%%%%%%%%%%%%%%%%%%%%%%%%%%%%%%%%%%%%%%%%%%%%%
\clearpage
\section{Data-related uncertainty on the generalization dataset}
\label{appendix_5}
This appendix details the data-related uncertainty obtained on the generalization dataset. The method used to quantify it is the Test-Time Augmentation technique. As mentioned in Section~\ref{section_5}, the standard deviation estimated on the predictions is on average equal to $1.16$ dB along $4,000$ km, against $3.22$ dB on the PE simulations. This indicates that the uncertainty induced on the TL by perturbing the initial $c_\text{ratio}$ slices with ten GW perturbation fields is not fully captured by $F_\theta(A_{z,d},f)$. However, even if the model acts as a low-pass filter, the data uncertainty associated with the predictions follows the global increase of uncertainty observed on the PE simulations in the absence of stratospheric waveguides (upwind scenarios). 

\smallskip

\begin{figure}
    \centering
    \includegraphics[width=6.65in]{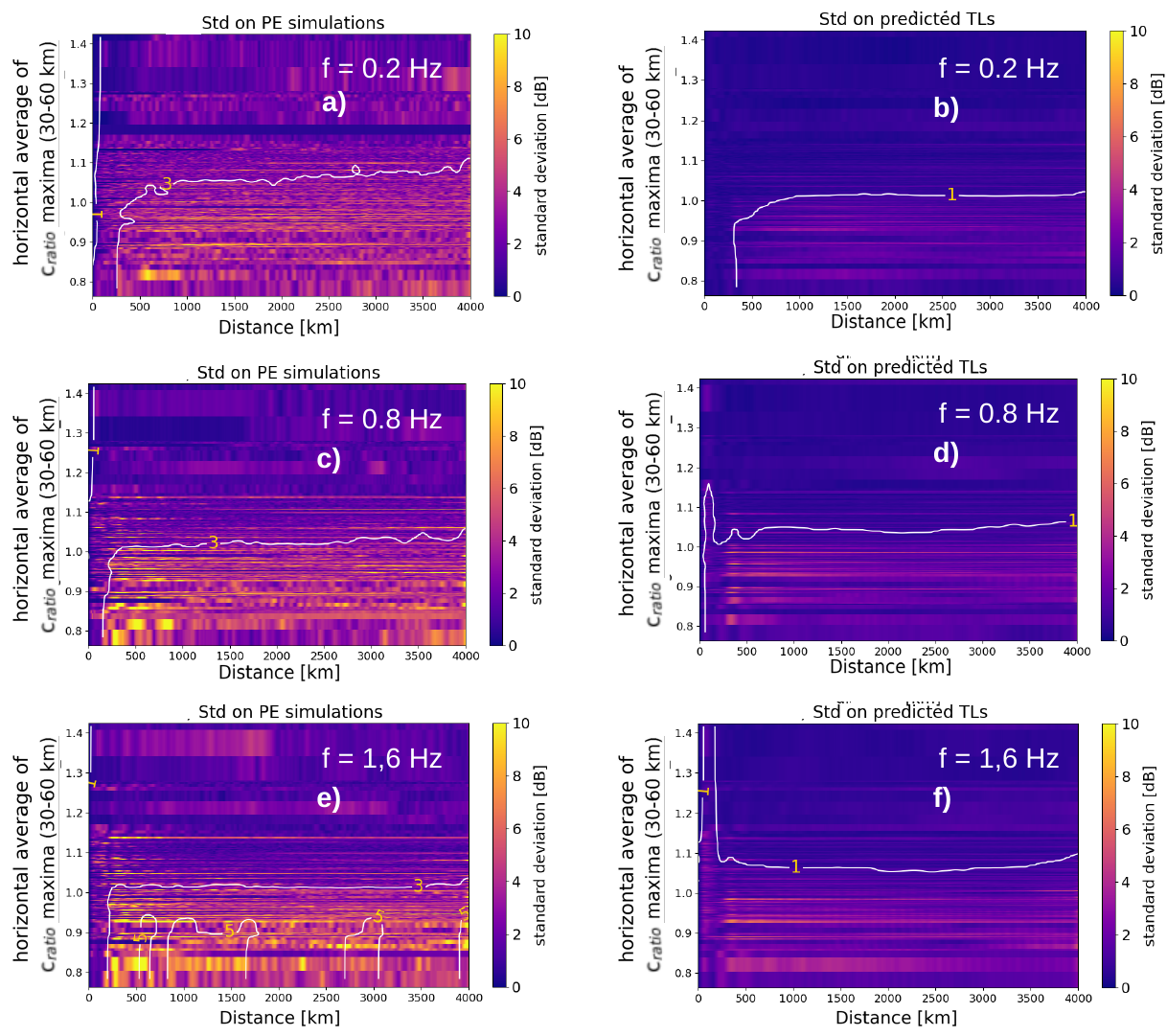}
    \caption{Uncertainty related to the input data, quantified on the generalization dataset using the Test-Time Augmentation technique.} 
    \label{fig:18} 
\end{figure}

Figure \ref{fig:18} shows that there is a slight increase in the average data-related uncertainty with increasing frequency; from $1$ dB at $0.1$ Hz to $1.5$ dB at $1.6$ Hz. Even if such a pattern is not visible in the PE simulations, we can link it with the increase of predictive errors with increasing frequency (see Section~\ref{results_test}).

%%%%%%%%%%%%%%%%%%%%%%%%%%%%%%%%%%%%%%%%%%%%%%%
\clearpage
\section{Reduction of the impact of the source frequency on the performances: test on two new databases}
\label{appendix_6}
Two new databases are created to train the neural network and try to limit the impact of the source frequency on its performance. The first new database (\textit{Alt-set-1}) is a database where the input samples correspond only to the atmospheric slices $A_{z,d}$ and where the outputs gather the ground-level TLs of all the five considered frequencies in a single two-dimensional array. Therefore, the resulting database is five times smaller than the one presented so far. It also provides a more comprehensive view of the impact of the source frequency on the expected TLs. The second new database (\textit{Alt-datasets-2}) corresponds to five independent datasets, whose outputs are ground-level TLs at one single frequency. Here again, the size of each database is reduced by five. A linear relationship has been established between the learning computation time of the models and the size of the database. 

\smallskip

Regarding the performances, the neural networks trained on the dataset presented so far and on the two new databases reach a median MRAE on the testing data in the same order of magnitude. The surrogate model presented so far reaches $7$\% of median MRAE, against $8.1$\% for the one trained on \textit{Alt-set-1} and $8.7$\% for the one trained on \textit{Alt-set-2}. However, neither of the two new approaches limits the degradation of the performance with increasing frequency. In addition, the generalization capabilities of the model trained on \textit{Alt-set-2} drop, with a median MRAE of $17.3$\% on the generalization dataset against $8.5$\% only for the model presented so far. As such, the neural network trained on the \textit{Alt-set-2} has not been selected for further tests. Regarding the model trained on \textit{Alt-set-1}, its main shortcoming lies in its capacity to predict at a stroke several ground-level TLs at different frequencies from input data of the form $A_{z,d}$ only. The absence of the frequency as a second part of the input limits the generalization capabilities to new source frequencies. 

\smallskip

As a result, the networks trained on the two new databases show degraded overall performances and additional shortcomings, while not limiting the impact of the source frequency on their predictions.

%%%%%%%%%%%%%%%%%%%%%%%%%%%%%%%%%%%%%%%%%%%%%%%
\clearpage
\section{Details on the comparisons between the training and the \textit{Tonga} set}
\label{appendix_7}
This appendix aims to detail the comparison realized between the training dataset and the \textit{Tonga-set} already discussed in Section~\ref{results_Tonga}. The process used to obtain the distributions presented in Figure \ref{fig:19} is similar to the one described in Appendix \ref{fig:15}. The results presented in this Appendix can be linked with Figure \ref{fig:12} which gives another point of view on the differences between the two datasets. 

\smallskip

The atmospheric slices of the \textit{Tonga-set} being extracted on a never-before-encountered location on Earth and on a new date, we observe strong differences between their distributions and the one obtained on the training data. These differences are visible for all the minimal, average, and maximal $c_\text{ratio}$ distributions for all considered atmospheric layers. Some departures correspond to over/under-represented atmospheric conditions in a given dataset, as previously observed when comparing the training and the generalization datasets (see Appendix \ref{fig:15}). For example, the \textit{Tonga-set} contains a strong over-representation of samples with a minimal $c_\text{ratio}$ of $0.75$ in the stratosphere or with an average $c_\text{ratio}$ of $1.04$ above $90$ km altitude. In addition to that, some distributions exhibit a horizontal shift compared to the training ones, such as the maximal distribution in the troposphere or the maximal one in the thermosphere. It is important to note that all these differences can interfere in the same atmospheric slice.

\begin{figure}
    \centering
    \includegraphics[width=6.5in]{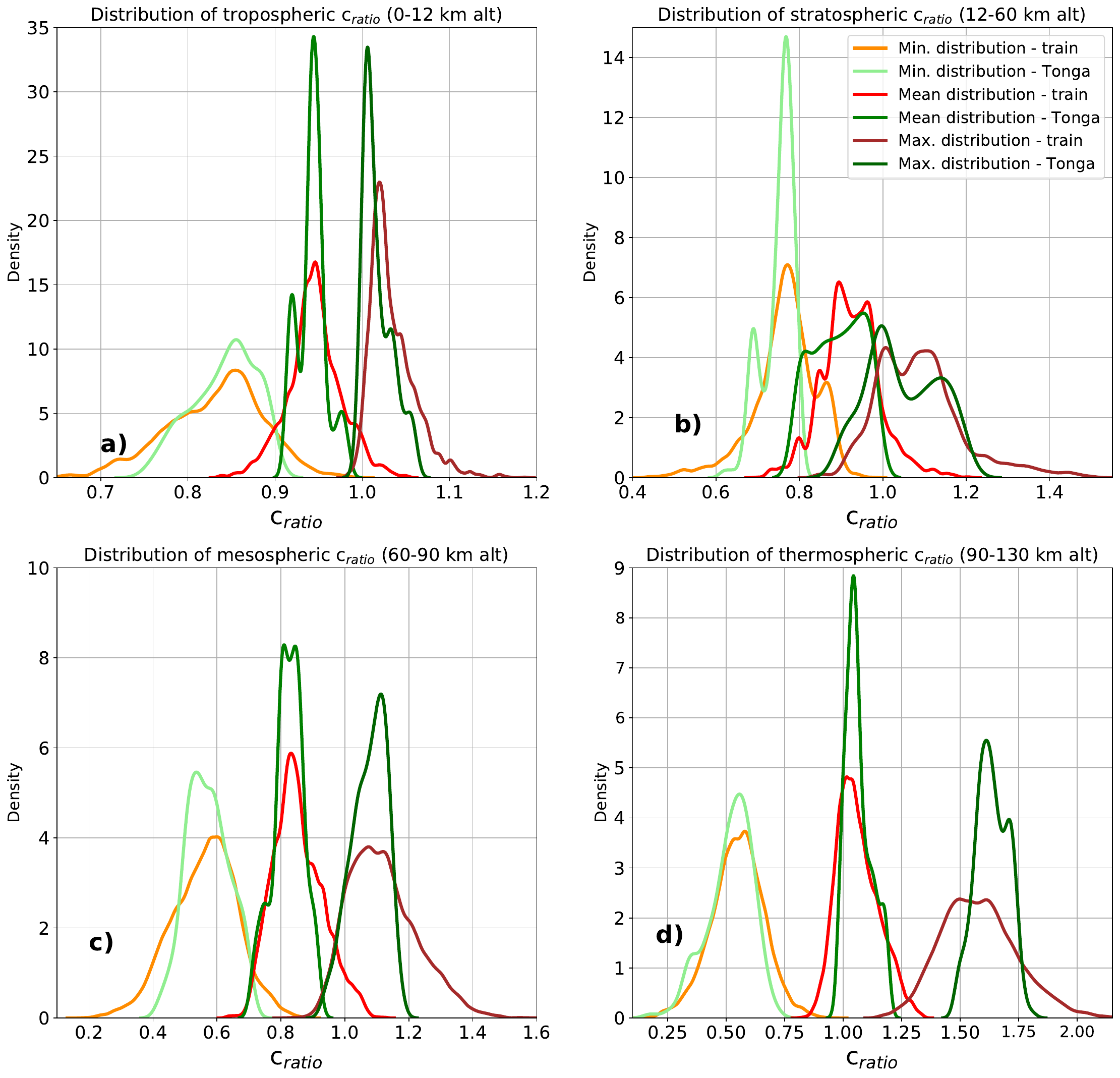}
    \caption{Distributions of minimal, average, and maximal values of $c_\text{ratio}$ in the troposphere, stratosphere, mesosphere, and thermosphere computed on the training dataset and the \textit{Tonga-set}.} 
    \label{fig:19} 
\end{figure}

%%%%%%%%%%%%%%%%%%%%%%%%%%%%%%%%%%%%%%%%%%%%%%%
\clearpage
\section{MRAE distribution according to the atmospheric conditions and source frequencies on the \textit{Tonga-set}}
\label{appendix_8}
\begin{figure}
    \centering
    \includegraphics[width=6.35in]{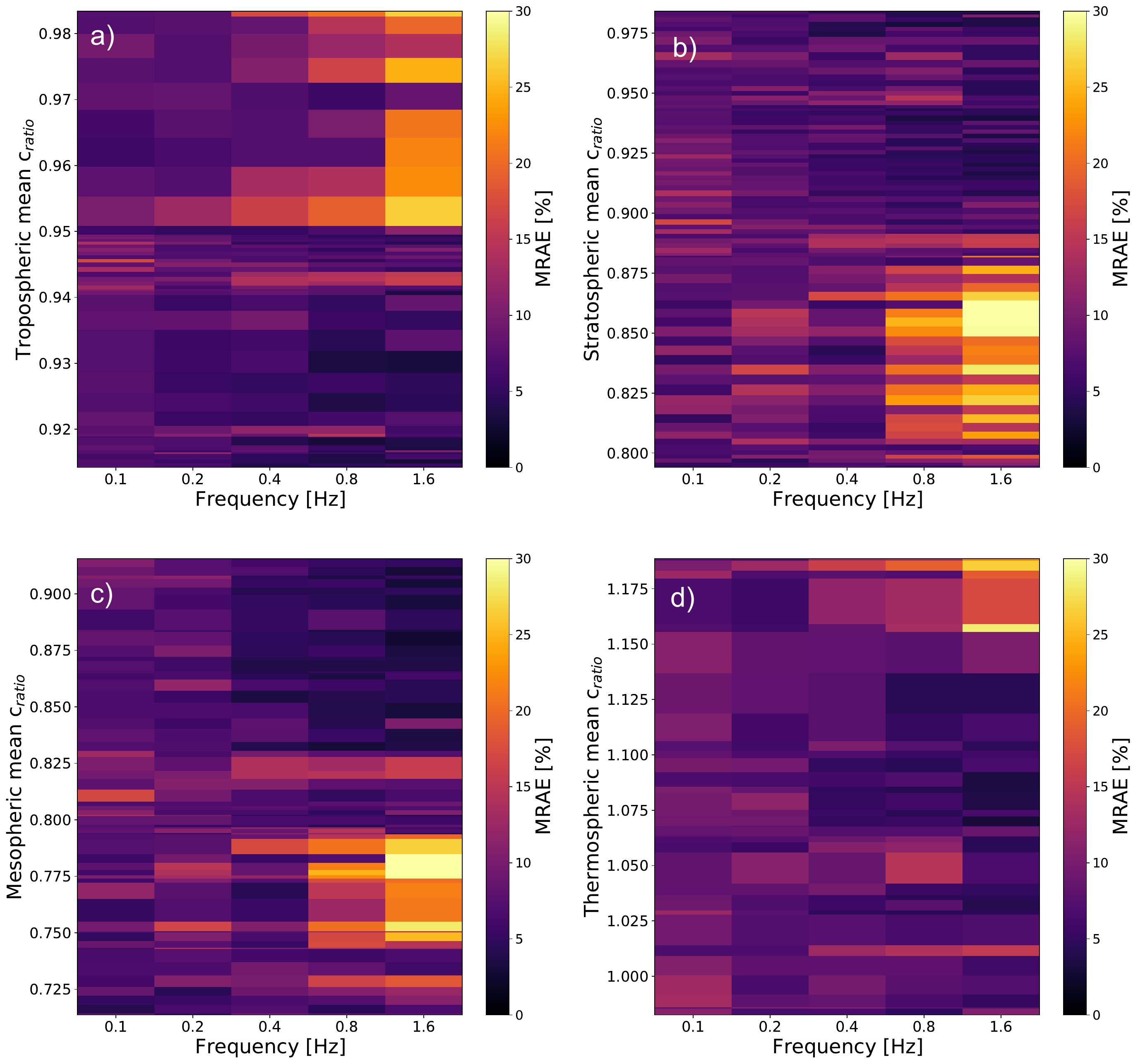}
    \caption{MRAE distribution according to the atmospheric conditions on the \textit{Tonga-set}, for the five usual source frequencies $f=0.1, 0.2, 0.4, 0.8, 1.6$ Hz.} 
    \label{fig:20} 
\end{figure}

The current Appendix shows the distribution of the MRAE according to the source frequency and the mean $c_\text{ratio}$ in the troposphere, stratosphere, mesosphere, and thermosphere on the \textit{Tonga-set}. The mean $c_\text{ratio}$ values are obtained by averaging the $c_\text{ratio}$ contained within the range of altitudes of each atmospheric layer and along the $4,000$ km-long propagation paths. Figure \ref{fig:20} highlights the areas associated with larger generalization errors ($\ge20$\%), particularly in the absence of stratospheric or mesospheric waveguides at $f\ge0.8$ Hz. Another region of higher errors is highlighted in panel a), for maximal mean $c_\text{ratio}$ values in the troposphere. We link this region with notable differences in the training dataset and the \textit{Tonga-set} atmospheric slices distributions (see horizontal shift of the \textit{Tonga} maximal distribution in the troposphere, Appendix \ref {fig:19}). 

%%%%%%%%%%%%%%%%%%%%%%%%%%%%%%%%%%%%%%%%%%%%%%%
\clearpage
\section{MRAE distribution according to the atmospheric conditions on the \textit{Tonga-set}, for five new source frequencies}
\label{appendix_9}
\begin{figure}
    \centering
    \includegraphics[width=6.35in]{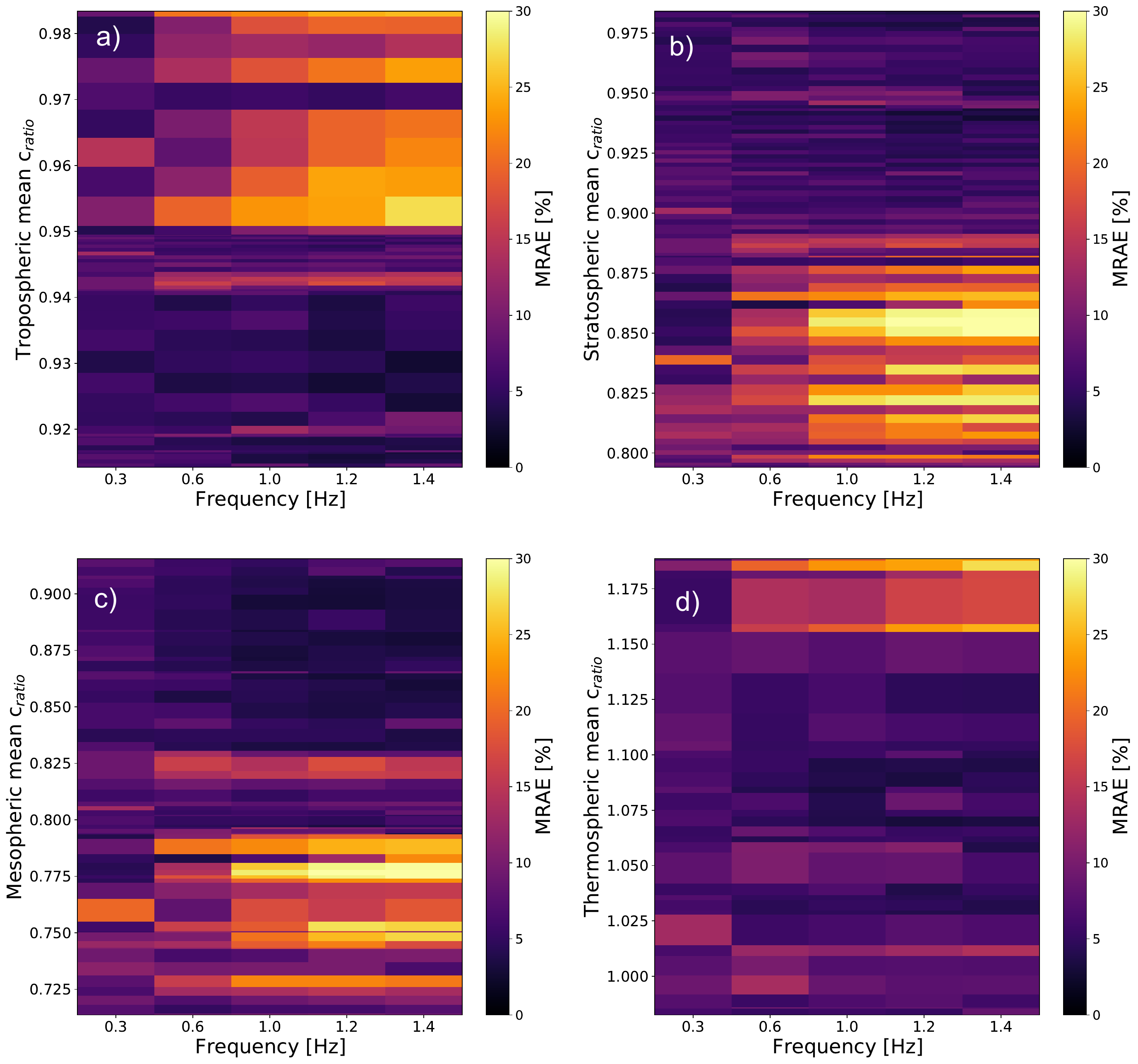}
    \caption{MRAE distribution according to the initial atmospheric conditions on the \textit{Tonga-set}, for five never-before-encountered source frequencies $f=0.3, 0.6,1.0, 1.2, 1.4$ Hz.} 
    \label{fig:21} 
\end{figure}

The current appendix analyzes further the generalization capabilities of the surrogate model by evaluating it on five never-before-encountered source frequencies: $0.3$, $0.6$, $1$, $1.2$ and $1.4$ Hz. Figure \ref{fig:21} details the distribution of the MRAE according to these new frequencies and the mean $c_\text{ratio}$ in the troposphere, the stratosphere, the mesosphere, and the thermosphere on the \textit{Tonga-set}. The figure highlights areas of larger generalization errors ($\ge20$\%), which can strongly be linked with the ones obtained by the model on the same atmospheric slices data, for the five usual source frequencies (see Appendix \ref{fig:20}). These regions with higher errors spread on the exact same mean $c_\text{ratio}$ values, but started as early as $0.6$ Hz.  

%%%%%%%%%%%%%%%%%%%%%%%%%%%%%%%%%%%%%%%%%%%%%%%
\clearpage
\section{Impact of the time coverage of the training database on the generalization capabilities}
\label{appendix_10}
We create an expanded database by sampling the Earth's atmosphere not only on January $15$, $2021$, but also on August $15$, $2021$. We build realistic atmospheric slices and run PE simulations following the processes described in Section \ref{section_2}. The neural network is trained on this expanded database as presented in Section \ref{section_3}. By doing so, we aim to investigate the effect of increasing the time resolution of the training database on the model's performance. The model presented in this study is referred to the "generic model" and the newly trained one as the "expanded model". 

\smallskip

The generic and the expanded models are both evaluated on a new generalization dataset composed of $360$ atmospheric slices around the military site of Hukkakero (Finland), extracted on August $20$, $2019$ and August $14$, $2020$. This site was selected because of its repeated explosions during a summer month in the northern hemisphere and is a benchmark event in the infrasound community \cite{ev23}.

\begin{figure}
    \centering
    \includegraphics[width=4.75in]{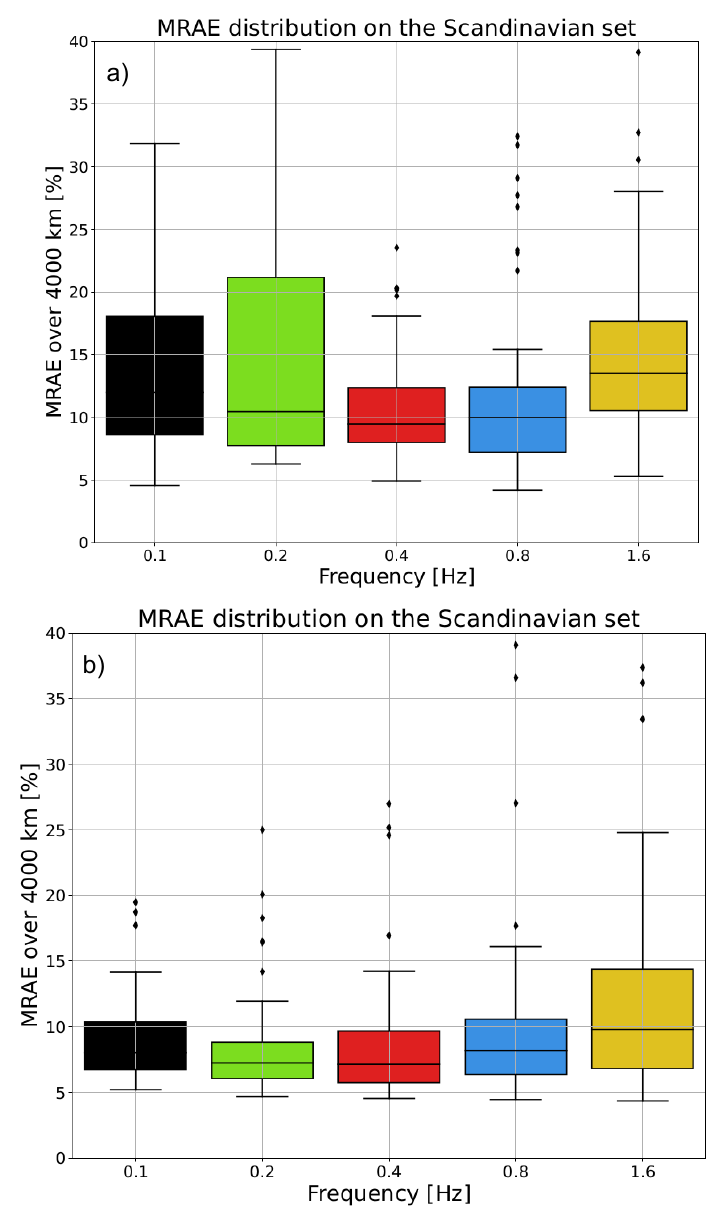}
    \caption{MRAE distributions along the $4,000$ km-long propagation paths for the five source frequencies, obtained by applying the generic model on the new generalization dataset (panel a)) as well as the expanded one (panel b)).} 
    \label{fig:22} 
\end{figure}

Figure \ref{fig:22} panel a) shows five boxplots (for the five source frequencies) summarizing the MRAE distributions along the $4,000$ km-long propagation paths obtained by applying the generic model on this new generalization dataset. Compared with the previous evaluations, a degradation of the performance is noted. For all source frequencies, the median error is around $10$ \%. We attribute this to the dates of August $14$, $2019$ and $20$, $2020$ being well outside the training domain of the generic model. On the other side, the expanded model shows improved performance (see panel b), with a significant reduction in median and quantile errors, especially below $0.4$ Hz. This seems to indicate that increasing the time resolution of the training database, by sampling the Earth at multiple dates (days, months, years), leads to better generalization capabilities of the neural network.

\clearpage

%%%%%%%%%%%%%%%%%%%%%%%%%%%%%%%%%%%%%%%%%%%%%%%
% DATA SECTION and ACKNOWLEDGMENTS
%%%%%%%%%%%%%%%%%%%%%%%%%%%%%%%%%%%%%%%%%%%%%%%
\section*{Open Research Section}
WACCM data (Atmospheric Chemistry Observations \& modeling, National Center for Atmospheric Research, University Corporation for Atmospheric Research; \citeA{agdata19}) were obtained via the NCAR Research data archive and are accessible at \url{https://doi.org/10.5065/G643-Z138}. The authors are grateful to the National Center for Physical Acoustics (NCPA) at the University of Mississippi for the PE modeling tool \textit{ePape} \cite{rwsoftware21} which is available on GitHub at \url{https://github.com/chetzer-ncpa/ncpaprop-release} or via \url{https://doi.org/10.5281/zenodo.5562713}. The TensorFlow library for Python \cite{masoftware15} can be downloaded from \url{https://doi.org/10.5281/zenodo.4724125}. \add[AJC]{The code to pre-process the input and output data, as well as testing the neural network is available on GitHub at} \url{https://github.com/Alice-Cameijo/DeepLearning_Infrasound} or via \url{https://doi.org/10.5281/zenodo.15319091}

\acknowledgments
The authors are grateful for the valuable insight provided by Pierre Andraud, PhD candidate at the French Atomic Energy Commission, and Andreas Steinberg, research scientist at the German Federal Institute for Geosciences in Hannover, on the uncertainty quantification in the field of deep learning. The authors would also like to thank Edouard Forestier, engineer at Naval Group, and the team of researchers at NORSAR for their previous works on the use of convolutional and recurrent neural networks for TL estimations. \add[AJC]{We are grateful to Sarah Albert and Pierrick Mialle for their careful reading of the manuscript and their valuable comments, which allowed to improve the paper.}

\smallskip

We would like to thank the Defense Innovation Agency for its financial support which made this project possible. Quentin Brissaud, geophysicist and data-scientist at NORSAR, and Sven Peter Näsholm, Associate Professor at the University of Oslo and senior research geophysicist at NORSAR, acknowledge support from the Research Council of Norway basic research program FRIPRO through the project \emph{Airborne inversion of Rayleigh waves} under Contract 335903. This study was facilitated by previous research realized within the framework of the ARISE and ARISE2 projects (\citeA{eb18}; \citeA{eb19blanc}), funded by the European Commission FP7 and Horizon 2020 programmes (grant nos. 284387 and 653980).

%%%%%%%%%%%%%%%%%%%%%%%%%%%%%%%%%%%%%%%%%%%%%%%
% REFERENCES and BIBLIOGRAPHY
%%%%%%%%%%%%%%%%%%%%%%%%%%%%%%%%%%%%%%%%%%%%%%%
\bibliography{references}

\end{document}